\documentclass[preprint,3p,twocolumn]{elsarticle}
\pdfoutput=1 
\usepackage{lineno,hyperref}
\usepackage{color}
\usepackage{amsmath}
\usepackage{xfrac}
\usepackage{subcaption}
\usepackage{verbatim}
\usepackage[percent]{overpic}
\modulolinenumbers[5]
\journal{NIMA}

\begin{document}

\begin{frontmatter}

\title{Conceptual design and first results for a neutron detector with interaction localization capabilities}

\author[mymainaddress]{J.~Heideman\corref{mycorrespondingauthor}}
\cortext[mycorrespondingauthor]{Corresponding author}
\ead{jheidema@vols.utk.edu}
\author[mymainaddress]{D.~P\'erez-Loureiro}

\author[mymainaddress,ORNLaddress]{R.~Grzywacz}
\ead{rgrzywac@utk.edu}
\author[mymainaddress]{C.R.~Thornsberry}
\author[mymainaddress]{J.~Chan}
\author[UTKNEaddress]{L.H.~Heilbronn}
\author[mymainaddress]{S.K.~Neupane}
\author[mymainaddress]{K.~Schmitt\fnref{now}}
 \fntext[now]{Present address: Los Alamos National Laboratory, Los Alamos, New Mexico 87545, USA}

\author[TTUaddress]{M.M.~Rajabali}
\author[TTUaddress]{A.R.~Engelhardt}
\author[TTUaddress]{C.W.~Howell}
\author[TTUaddress]{L.D.~Mostella}
\author[TTUaddress]{J.S.~Owens}
\author[TTUaddress]{S.C.~Shadrick}
\author[UKaddress]{E.E.~Peters}
\author[UKaddress]{A.P.D.~Ramirez}
\author[UKaddress]{S.W.~Yates}
\author[Agileaddress]{K. Vaigneur}


\address[mymainaddress]{Department of Physics and Astronomy, University of Tennessee, Knoxville, Tennessee 37996 USA}
\address[UTKNEaddress]{Department of Nuclear Engineering, University of Tennessee, Knoxville, Tennessee 37996 USA}
\address[ORNLaddress]{Physics Division, Oak Ridge National Laboratory, Oak Ridge, Tennessee 37831 USA}
\address[TTUaddress]{Department of Physics Tennessee Technological University, Cookeville, Tennessee, 38505, USA}
\address[JINPAaddress]{Joint Institute for Nuclear Physics and Applications, Oak Ridge, Tennessee 37831 USA}
\address[UKaddress]{Departments of Chemistry and Physics \& Astronomy, University of Kentucky, Lexington, Kentucky, 40506 USA}
\address[Agileaddress]{Agile Technologies, Inc., Knoxville, Tennessee 37920 USA}

\begin{abstract}
A new high-precision detector for studying neutrons from beta-delayed neutron emission and direct reaction studies is proposed. The Neutron dEtector with Xn Tracking (NEXT) array is designed to maintain high intrinsic neutron detection efficiency while reducing uncertainties in neutron energy measurements. A single NEXT module is composed of thin segments of plastic scintillator, each optically separated, capable of neutron-gamma discrimination. Each segmented module is coupled to position sensitive photodetectors enabling the high-precision determination of neutron time of arrival and interaction position within the active volume. A design study has been conducted based on simulations and experimental tests leading to the construction of prototype units. First results from measurements using a $^{252}$Cf neutron source and accelerator-produced monoenergetic neutrons are presented. 
\end{abstract}


\end{frontmatter}


\section{Introduction}
The new generation of radioactive ion-beam facilities will enable access to very neutron rich nuclei, approaching, and even reaching the neutron drip-line in certain cases \cite{FRIB}.
Far from stability, neutron separation energies decrease as beta-decay endpoint energies become large, increasing the likelihood of beta-delayed neutron emission. In these regions, neutron spectroscopy becomes essential to obtain important information about the nuclear structure of these nuclei \cite{BLANK2008403,RevModPhys.84.567,NAKAMURA201753}. 
The Neutron dEtector with Xn Tracking (NEXT) array has been developed to observe beta-delayed neutron emitters with improved precision. 
These improvements will also be applicable to proton transfer reactions which probe discrete states of exotic nuclei.

NEXT has also been designed to incorporate neutron-gamma (n-$\gamma$) discrimination to improve background rejection. Prevalent gamma-ray backgrounds common in decay and reaction measurements can be reduced significantly in neutron spectra, further improving neutron energy determinations \cite{FEBBRARO2018189}. Proof-of-principle tests for the NEXT design will be described along with results from the first neutron measurements.

\section{Detector Design}

 A neutron time-of-flight (ToF) detector determines neutron kinetic energies, $E$, by measuring the time difference, $T$, between a START signal (associated with the initial emission of a neutron) and a STOP signal (detection of emitted neutron) over some fixed distance, $L$. In such a configuration, a simple calculation for the energy resolution, $\Delta E$, as a function of detector limitations is given by the following expression \cite{KORNILOV2009226}:
\begin{equation}
\frac{\Delta E}{E}=\sqrt{\left(\frac{2\Delta T}{T}\right)^2+\left(\frac{2\Delta L}{L}\right)^2},
\label{eq:resolution}
\end{equation}
in which $\Delta T$ is uncertainty in the time-of-flight, $T$, of the neutron and $\Delta L$ is the uncertainty in neutron flight-path length, $L$. Therefore, the energy resolution is directly related to the timing resolution of the detection system and the precision in the measurement of the path length. The latter is mainly due to the uncertainty in the determination of the interaction point within the detector. Thick detectors maintain good neutron detection efficiency at the expense of position resolution. Typical thicknesses for plastic-scintillator-based neutron ToF detectors are about 2-3~cm, which is a good trade-off between position resolution and efficiency \cite{MORRISSEY1997222,BUTA2000412,PETERS2016122}.
\begin{figure}[tp]
\centering
\includegraphics[width=0.5\textwidth]{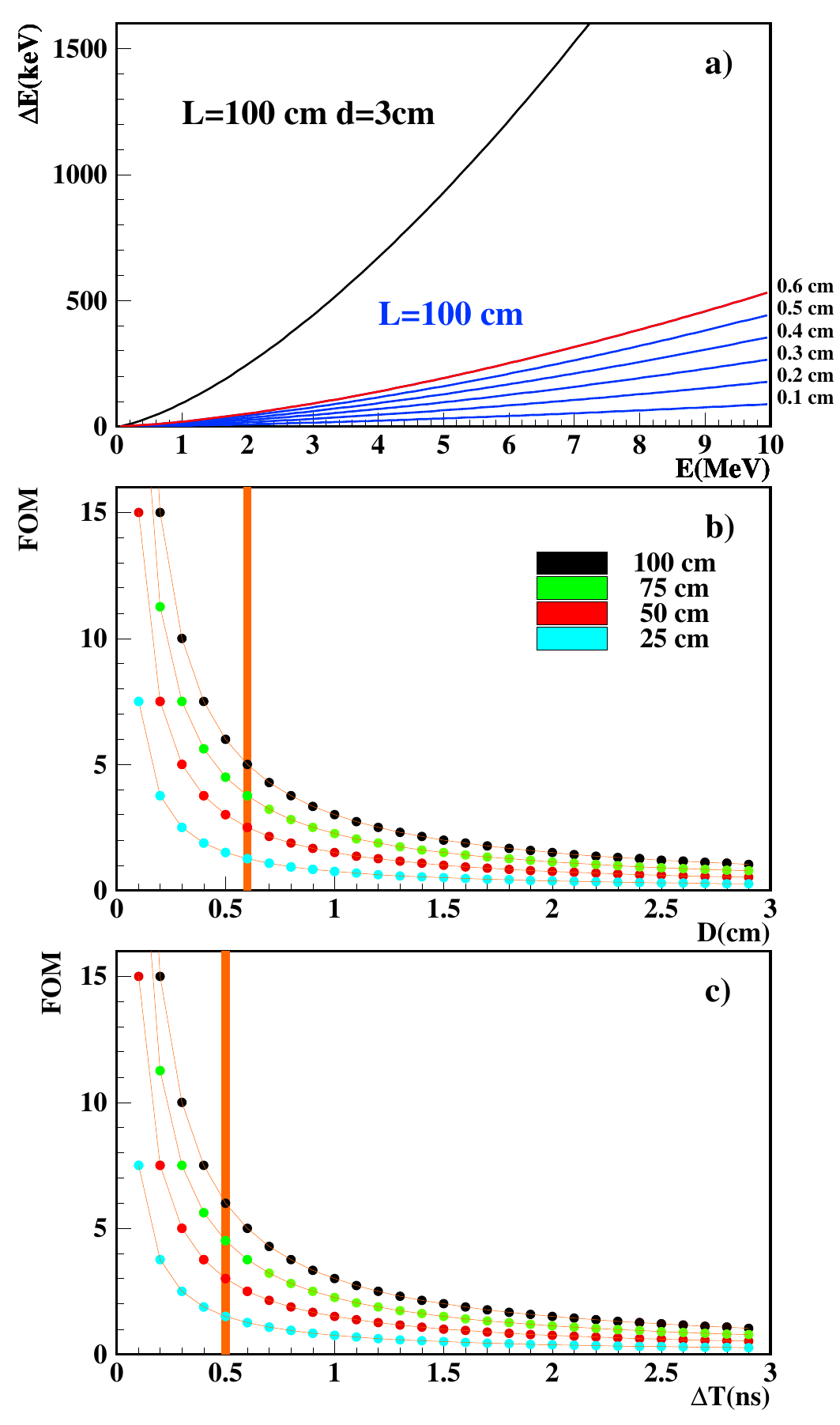}
\caption{(a) Energy resolution of varying prototype tile thicknesses (1-6~mm) as a function of neutron energy. The black curve represents current energy resolution calculations as a function of energy for VANDLE ($L$=100~cm, $\Delta L$=3~cm, and $\Delta T$=3~ns). (b) and (c) show figures of merit calculations as a function of tile thickness, $D=\Delta L$, and ToF resolution, $\Delta T$, respectively, for 1~MeV neutrons. The colored data points correspond to different flight-path lengths. In (b), $\Delta T$ has been fixed at 500~ps and in (c), $D$ has been fixed at 6~mm. The orange vertical line in (b) and (c) are the expected detector performance with current proposed design specifications ($\Delta L$=6~mm and $\Delta T$=500~ps).}
\label{fig:resolution}
\end{figure}
Detailed simulations will be presented in Sect. \ref{sec:Simulations} to show how detector thickness and photon propagation affect ToF resolution.

\subsection{Detector Requirements}

The design of the NEXT module is dictated by the necessity to obtain the highest possible precision in measurements of position and timing of the neutron, consistent with the capabilities of the constituent scintillator material and sensors. The optimal, realistic segmentation must warrant sufficiently good light collection in order to retain timing and n-$\gamma$ discrimination capabilities. Fig. \ref{fig:resolution} shows the calculated dependencies of the neutron energy resolution on the thickness and timing resolution resulting using Eq. \ref{eq:resolution} for a 100~cm neutron flight path. As a comparison to a state-of-the-art neutron ToF detector, the black curve in Fig. \ref{fig:resolution}(a) represents typical VANDLE capabilities \cite{PETERS2016122} and shows that the uncertainty in the neutron path distance is the main limitation in the energy resolution of this detector, especially at neutron energies above 2~MeV. Figs. \ref{fig:resolution}(b) and \ref{fig:resolution}(c) show the expected figure of merit $\left(FoM = \frac{\Delta E_{VANDLE}}{\Delta E}\right)$ as a function of position resolution ($D$) and timing resolution ($\Delta T$), respectively, for 1~MeV neutrons. The colored data points represent different neutron path lengths, $L$, with the orange vertical lines in each plot corresponding to NEXT design goals.

In a ToF setup with a given $\frac{\Delta L}{L}$, the $\frac{\Delta T}{T}$ should be commensurate, i.e., for a particular timing resolution $\Delta T$ and flight path length $L$, there is a minimal segment thickness $\Delta L$ beyond which the timing resolution dominates the overall energy resolution. Detectors with $\Delta T$=1~ns ToF resolution at distances $L$=50-100~cm from the decay or reaction should be no more than 10~mm thick. Thin detectors (3-6~mm) will have a smaller contribution to the overall uncertainty from the determination of the interaction position, amounting to a larger contribution from the ToF resolution.

\begin{figure}[tp]
  \centering
  \includegraphics[width=\linewidth, trim={3cm 3cm 3cm 1cm}, clip]{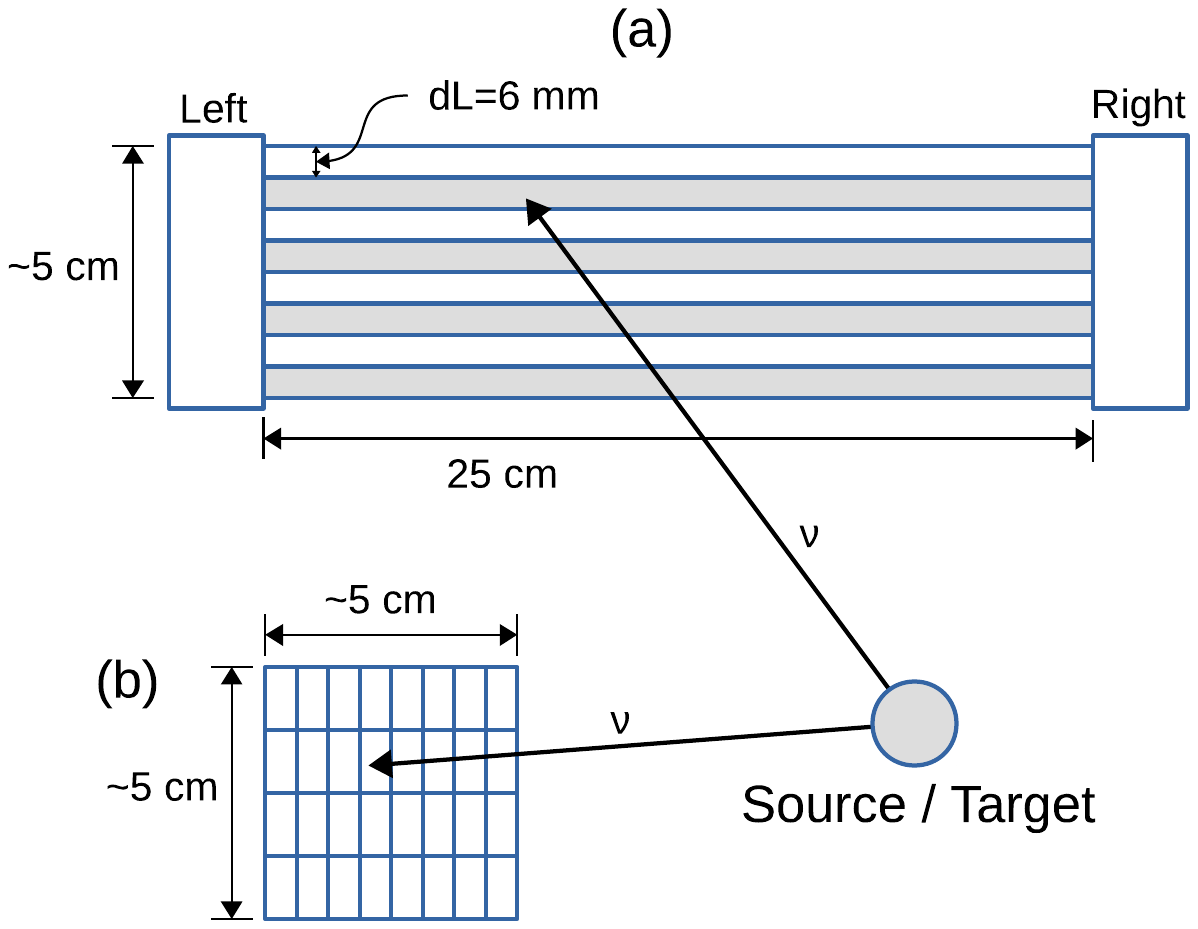}
  \caption{Schematic showing a potential segmentation configuration for a single NEXT module in a time-of-flight setup. (a) A top-down view of the detector segmentation along the neutron flight path. (b) Side view of a prototype showing the individual segments within the detector, each optically separated from one another. Vertically aligned segments in (b) are designated as columns and horizontally aligned segments are designated as rows.}
  \label{fig:NEXTschematic}
\end{figure}
Calculating neutron energies from ToF measurements can also benefit if neutrons are discriminated from gamma rays. Scattered gamma rays present during an experiment can cause significant background in the neutron time-of-flight spectrum.
There are several commercially available solid-state scintillators which can provide strong n-$\gamma$ discrimination and timing, such as stilbene, anthracene, and para-terphenyl. Due to the limited ability of machining these scintillators into large segmented arrays, these are presently not preferable or economically feasible as a material for the proposed detector. When a viable n-$\gamma$ discriminating plastic scintillator, Eljen 299 (EJ-299) \cite{eljen}, was first developed \cite{Zaitseva2012}, the material was not hard enough to facilitate machining and had to be cast directly into the final detector geometry. However, recent improvements to the EJ-299 polymer matrix increased physical stability and pulse-shape discrimination (PSD) capabilities \cite{ZAITSEVA201897}. This new plastic scintillator, Eljen 276 (EJ-276), has n-$\gamma$ discrimination capabilities comparable to liquid scintillators and is now firm enough for machining of segments with appropriate thickness and geometry to construct a high-resolution neutron time-of-flight detector.

\subsection{Detector Concept}

\begin{figure}[t]
 \centering
 \begin{subfigure}{\columnwidth}
  \centering
  \includegraphics[width=\linewidth,trim={3.7cm 0 0 0},clip]{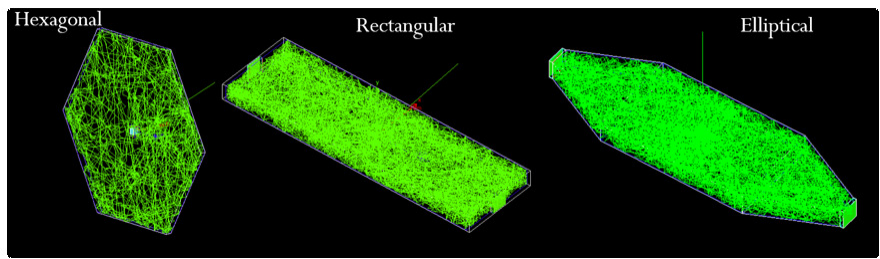}
 \caption{}
 \end{subfigure}%
 \\
 \begin{subfigure}{\columnwidth}
  \centering
  \includegraphics[width=\linewidth]{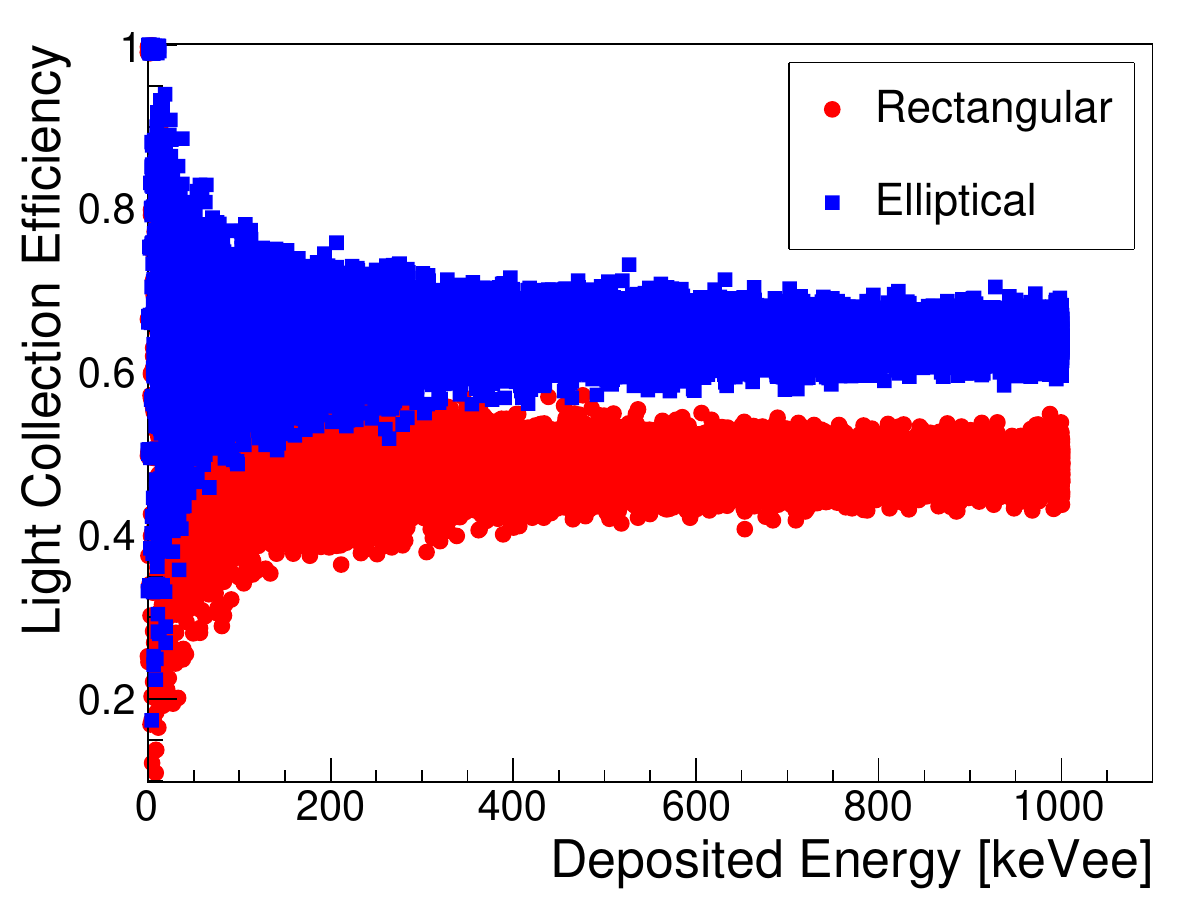}
  \caption{}
 \end{subfigure}%
 \caption{(a) Visualization of a 1~MeV neutron event in NEXT\emph{sim} for two Mylar-wrapped geometries considered for a NEXT layer with single photosensors on either end. Green lines correspond to optical photons produced in the scintillation. (b) The light collection efficiency for the two geometries when observing 1~MeV neutrons. The detector in the elliptical geometry observes, on average, 68\% of all photons produced in these events compared to 50\% in the rectangular geometry.}
 \label{fig:GeoEff}
\end{figure} 

The NEXT array is based on multi-layered modules of PSD plastic scintillator with position sensitive photodetectors on both ends of a single module. Each detector consists of eight $\sim$6-mm-thick layers, with an approximate effective thickness of 4.8~cm. These dimensions were driven, among other factors, by the availability of small form factor photosensors and will be substantiated later in Sects. \ref{sec:SiPMs} and \ref{sec:EJ276}. Fig. \ref{fig:NEXTschematic} shows a possible multi-layered module configuration, with segmentation along the horizontal and vertical directions, wherein the vertically aligned segments with respect to the incoming particle are denoted as columns or layers and the horizontally aligned segments along the incoming particle trajectory are rows. The best timing and position resolution are achieved by orienting the higher segmentation along the direction of incident particles. The photosensors considered are either an array of silicon photomultipliers (SiPMs) or flat panel multi-anode photomultiplier tubes (MAPMTs). To reduce the number of acquisition channels, an Anger Logic readout will be used in conjunction with the photosensors \cite{ANGER}. Analysis of the position-sensitive Anger Logic response from detected scintillation light will determine the specific layer in which the neutron scattered, reducing the uncertainty in the neutron flight-path length, $\Delta L$. The fast-timing (sub-nanosecond) capabilities of these photosensors will further improve energy resolution by reducing the ToF uncertainty, $\Delta T$.

\section{NEXT Simulations} \label{sec:Simulations}
In order to investigate the light collection efficiency as well as the timing capabilities of the different layer geometries, NEXT\emph{sim}, a {\sc Geant4}-based code was developed \cite{AGOSTINELLI2003250,ALLISON2016186}. The NEXT\textit{sim} code uses {\sc Geant4} version 10.1 Patch 3 and outputs to {\sc Root} files for further analysis.
The software simulates the interaction of neutrons, gammas, and charged particles in the matter they traverse. The physics model (referred to as \emph{Physics List} in the {\sc Geant4} context) employed is the recommended QGS\_BERT\_HP, which includes the standard electromagnetic and high-precision models for neutron scattering, elastic and inelastic, as well as capture and fission. This model is based on the G4NDL evaluated neutron data library \cite{Apostolakis2009}.

Neutron-induced scintillation is simulated using the associated light response from energy deposited in the scintillator by a scattered neutron. This relationship for organic scintillators is detailed in \cite{VERBINSKI19688} and has been scaled appropriately to the light output and scintillation efficiency for EJ-276. The G4OpticalPhysicsList is included to treat the transport of each photon until it escapes the active volume, is absorbed, or is detected. For each event, the position and timing information for each neutron scatter within the detector volume is recorded along with relevant photon information, such as the minimum and average photon arrival time along with the position information of all detected photons at the photosensor surface.  

 Different geometries and wrappings considered for NEXT modules displayed in Figure~\ref{fig:GeoEff}(a) can be generated, e.g., rectangular bars and elliptical bars (bars in which the corners were cut in angle to maximize the light focusing in the detector). The ends of each scintillator segment are coupled to photosensitive surfaces with a thin layer of optical grease. Any of the available geometries, scintillator, and wrapping materials can be chosen via macro-driven commands.

\begin{figure}[tb]
\centering
\includegraphics[width=\linewidth]{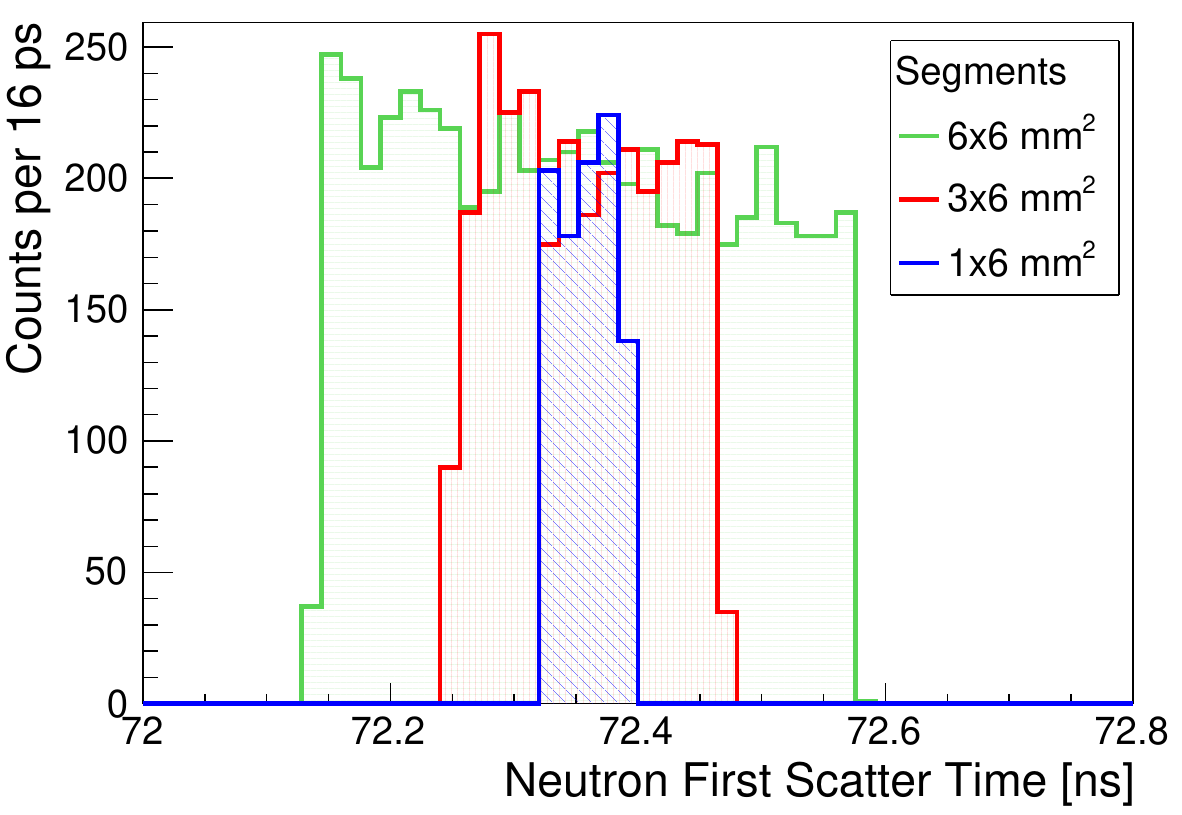}
\caption{Distribution of the time-of-flight for a pencil beam of 1~MeV neutrons impinging onto scintillator bars of varying thickness at a distance of 1~m. ToF was calculated as the neutron's first scattering time in the scintillator.}
\label{fig:simToF}
\end{figure}

\subsection{Simulation of light collection efficiency}

The different 6-mm-thick geometries considered for single NEXT layers shown in Fig. \ref{fig:GeoEff}(a) were studied using NEXT\emph{sim} in order to determine the light collection efficiency when wrapped in Mylar and coupled to two photosensors, one on each end. Fig. \ref{fig:GeoEff}(b) shows the results of the light collection efficiency, calculated as the ratio between the number of detected and produced photons, as a function of the energy deposited in the scintillator. In this case, the photosensors are considered ideal meaning every photon hitting the sensitive surface will be detected. The average efficiency of the rectangular geometry is 50\%. The average efficiency of the elliptical geometry is higher and reaches 68\% due to the focusing effect towards the photosensors. Elliptical geometries were originally considered for single-sensor readouts but further tests were not conducted after it was decided that NEXT must be comprised of multiple layers.

\begin{figure}[tb]
\centering
\includegraphics[width=\linewidth, trim = {0 0 0 0}, clip]{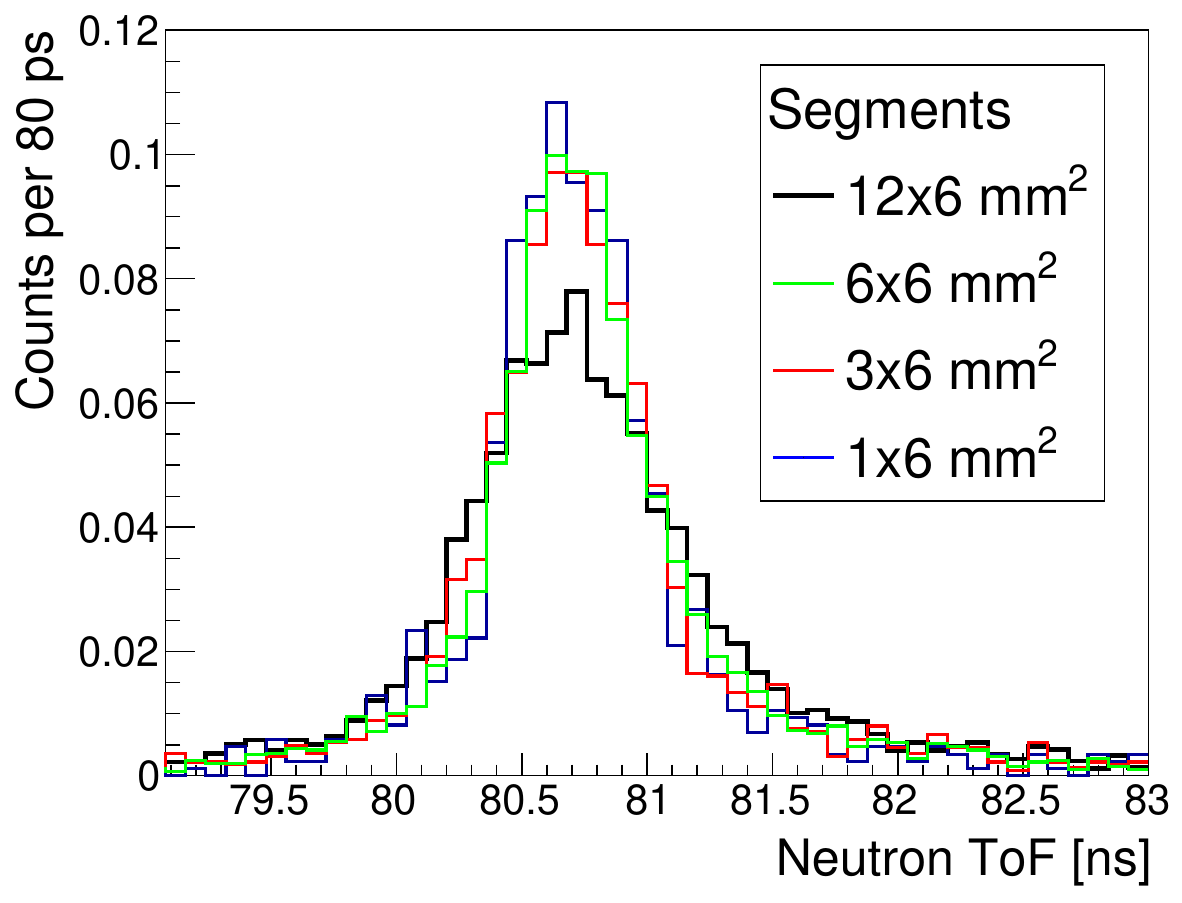}
\caption{Distribution of the time-of-flight for a pencil beam of 1~MeV neutrons scattering in scintillator plates 1~m away. The PolyCFD analysis \cite{PhDCory} was applied to both photosensor responses for each event to calculate high resolution times in order to compute the neutron flight times. Normalized ToF distributions are shown for 10~cm long plates with geometric cross-sections of $1 \times 6$~mm$^{2}$ (blue), $3 \times 6$~mm$^{2}$ (red), $6 \times 6$~mm$^{2}$ (green), and $12\times6$~mm$^2$ (black).}
\label{fig:plateTOF}
\end{figure}

\subsection{Simulation of photosensor response} \label{sec:photosensorResponse}

The photosensor response to the detected scintillation light was also added to the simulations for better comparisons to experimental timing tests. A Single Photo-Electron (SPE) response function, specific to each photosensor, is folded with the optical-photon arrival-time distribution to obtain realistic photomultiplier signals. The SPE response functions for SiPMs and PMTs were taken from \cite{Choong2009}. The total response is the sum of the SPEs of each photon detected by the photosensor weighted by the product of the anode gain and quantum efficiency. The resultant light-response pulse is given a baseline offset and electrical noise, and is then ``digitized" by placing it into discrete bins on the y-axis (e.g., from 0 to 65535 to represent a 16-bit digitizer) and discrete time bins on the x-axis (e.g., 4~ns for a 250~MSPS system). The digitized pulses are then integrated to obtain the representative light yield of the event and are processed with a polynomial constant fraction discrimination algorithm (PolyCFD) \cite{PhDCory} as discussed later in Sect. \ref{sec:HRT}. The PolyCFD algorithm computes a time for each pulse which represents the time-of-flight of the incident neutron aggregated from all collected photons. 

\subsection{Simulation of time-of-flight resolution of the detector}
Simulations of 1~MeV neutrons impinging onto 10-cm-long plastic scintillator bars were used to establish the timing resolution dependence on detector thickness. Geometrical cross-sections of $1 \times 6$~mm$^{2}$, $3 \times 6$~mm$^{2}$, and $6 \times 6$~mm$^{2}$ were studied. The neutron's first scattering time was used to directly measure the neutron ToF. This method provides the most direct information to indicate the timing uncertainty associated with the thickness of the detector. The ToF distributions can be seen in Fig. \ref{fig:simToF}, where an increase in the thickness of the scintillator bar results in a broadening of the ToF resolution. This effect is due to the uncertainty in the interaction position along the neutron path within the scintillator and the non-negligible flight time of the neutron.

To test the timing resolution of a digitized photosensor response when using bars of different thicknesses, scintillator plates with cross-sections $1 \times 6$~mm$^{2}$, $3 \times 6$~mm$^{2}$, $6 \times 6$~mm$^{2}$, and $12 \times 6$~mm$^{2}$ were modeled and the light output and timing of each plate response is computed as in Sect. \ref{sec:photosensorResponse}. The PolyCFD timing information from the previously simulated events in different thicknesses was used. The neutron ToF is computed as the average of PolyCFD HRT for the left and right photosensors. Fig. \ref{fig:plateTOF} shows the normalized ToF distributions for each plate overlaid on one another. The difference in ToF distributions between Figs. \ref{fig:simToF} and \ref{fig:plateTOF} is due to the contributions of the photon arrival time distribution accounted for in Fig. \ref{fig:plateTOF}. The FWHM ToF resolution for the 1~mm, 3~mm, and 6~mm thick plates are all within $7\%$ and average to 600~ps. The ToF resolution of the 12~mm plate was 54\% larger as compared to the average of the other plates. When a complete digitized analysis is considered, the ToF resolutions of the scintillator segments with varying thicknesses are dominated by the photon collection and trace analysis up to $\sim$6~mm thickness, beyond which the detector geometry begins to contribute to the ToF resolution. It was also concluded from simulations that the detection efficiency of a plate scales linearly with its thickness (i.e., the efficiency of the $6 \times 6$~mm$^2$ plate is two times larger than the $3 \times 6$~mm$^2$ plate). This result means that, in addition to exhibiting six times greater efficiency, the $6 \times 6$~mm$^2$ plate exhibits approximately the same detector time resolution as the $1 \times 6$~mm$^2$ plate when coupled to a realistic acquisition system.

\begin{figure}[tb]
\centering
\includegraphics[width=\columnwidth]{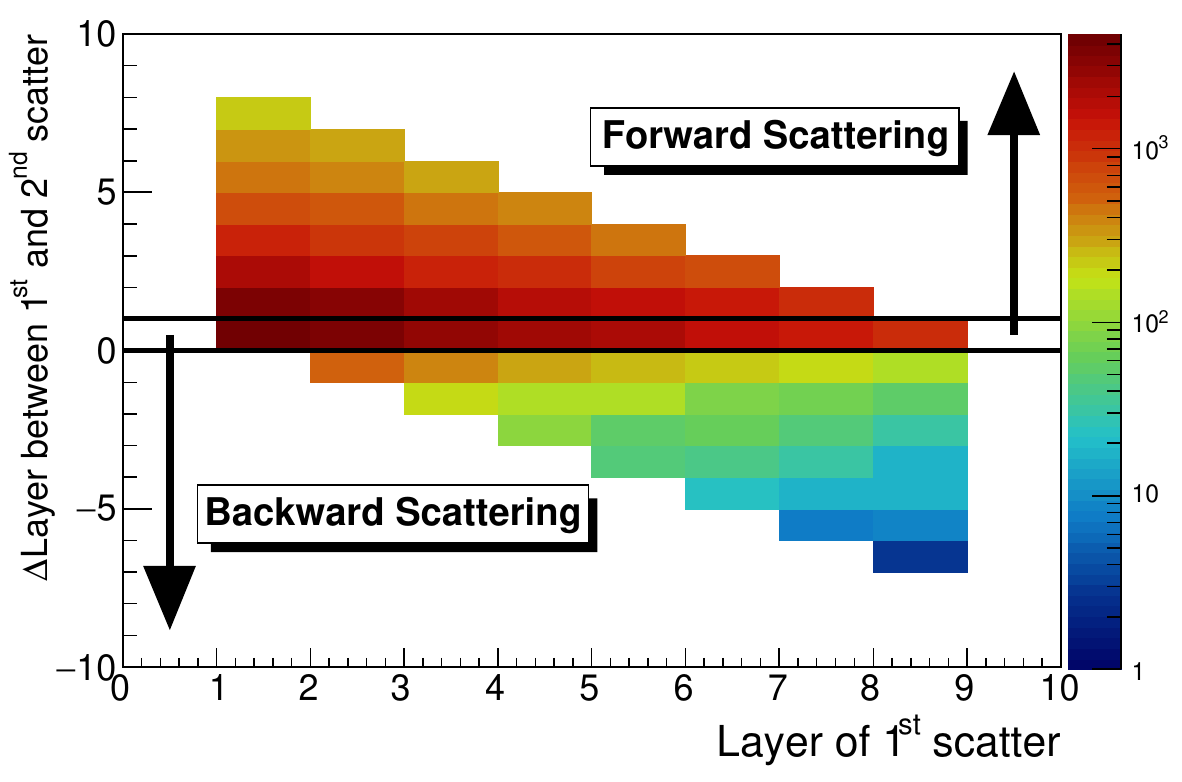}
\caption{Two-dimensional histogram showing 1~MeV neutron multiple scattering between layers of a NEXT detector. The x-axis denotes the layer number of the first neutron interaction and the y-axis denotes the difference in layer number between the first and second neutron interaction layer. Forward scattering events (above the two black lines) comprise about 57\% of events while backward scattering events (below black lines) are about 6.1\%. The remaining events (between black lines) are a result of scattering more than once in the same layer.}
\label{fig:Scattering}
\end{figure}

Based on the loss of ToF resolution for the 1~mm and 3~mm layers when using the full photosensor response, seen in Fig. \ref{fig:plateTOF}, it was decided that the minimum layer thickness of a prototype using EJ-276 plastic scintillator should not be less than 6~mm; thinner tiles would not provide any further benefit to ToF measurements due to the limited timing resolution of the data acquisition system.
 
\subsection{Study of neutron multiple scattering} \label{sec:nMulti}

If a neutron scatters multiple times within the detector, the neutron energy may be incorrectly determined. 
The \emph{NEXTsim} code was used to evaluate the probability and effects of multiple scattering events in the different columns (layers) of the detector. A multi-layer detector was modeled using 6$\times$12.7$\times$254~mm$^3$ scintillator cells arranged in a 4$\times$8 segmented detector, the same design as shown in Fig. \ref{fig:NEXTschematic}. An 8$\times$8 multi-anode photosensor (6$\times$6~mm$^2$ anodes), similar to commercially available designs, was coupled to each end. The simulation tracks the neutron while it traverses the entire detector. Fig. \ref{fig:Scattering} shows first and second scattering layer differences in the multi-layered detector obtained from the simulation of a 1~MeV neutron knife beam (uniformly distribution along a vertical line at the center between the two ends of the detector). Of all neutrons that interact within the detector, 69.5\% will scatter more than once in the scintillator. Relative to the first interaction point, the majority of multiple-scattered neutrons (57\%) forward scatter, while only 6.1\% of events scatter backwards. 
\begin{figure}
  \centering
  \includegraphics[width=\columnwidth]{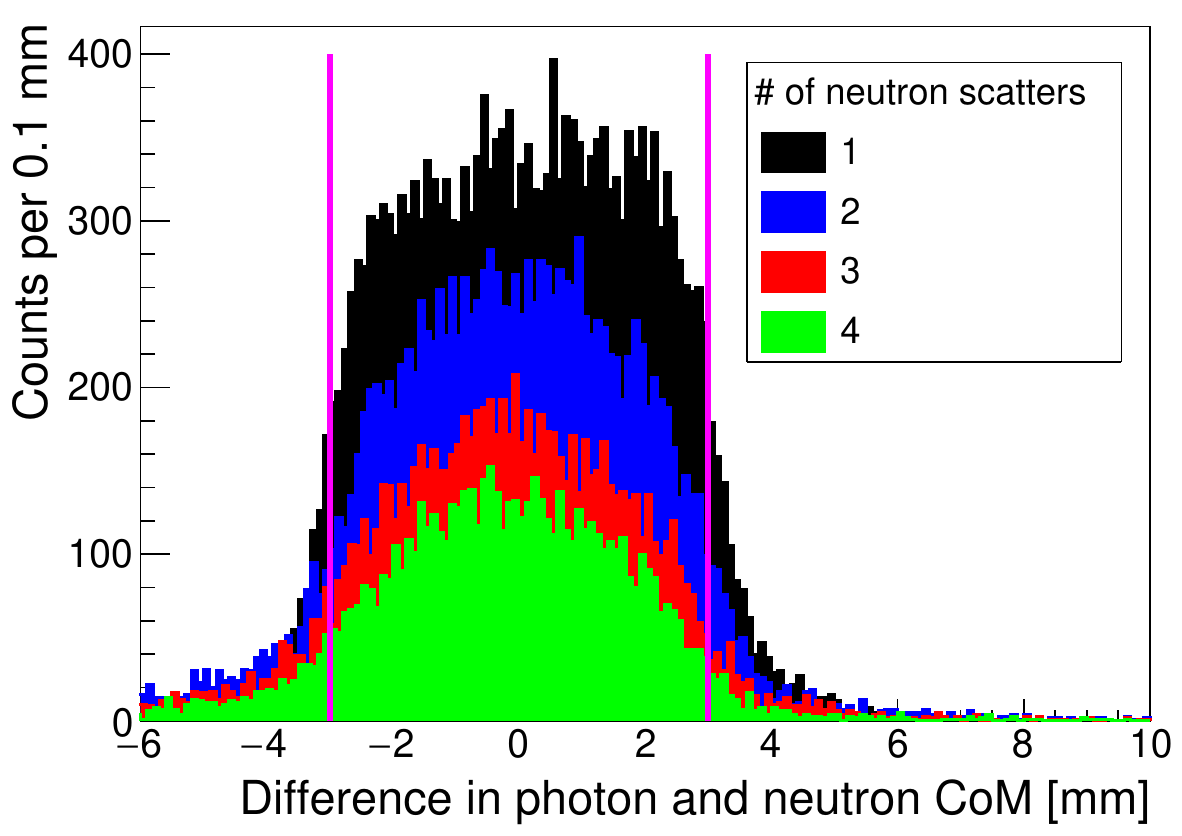}
  \caption{Differences between reconstructed photon center-of-mass scatter position and neutron center-of-mass position shown here for 1 (black), 2 (blue), 3 (red), and 4 (green) scatters of 1~MeV neutrons within the detector. The magenta lines represent the width of a single layer thickness. Events outside the lines would be considered improperly reconstructed.}
  \label{fig:PhotonNeutronPos}
\end{figure}

Neutron multiple scattering will have a large effect on the reconstructed scintillation position within the detector using a center-of-mass (CoM) analysis for detected photons. The optical photon CoM (segment position) is computed by taking the weighted average of the X and Y positions of all photons detected at the surface of the photosensor. Each detected photon position is weighted using the product of the gain of the anode at which it was detected and the quantum efficiency of the photosensor for a given wavelength. The photon CoM analysis is analogous to the Anger Logic position algorithm proposed for the NEXT prototype. Ideally, the reconstructed photon CoM should be within the same column as the scattered-neutron CoM, defined as the average X and Y interaction position weighted by the imparted energy for each scatter within the detector. From the same simulated 1~MeV neutron data analyzed to study the forward and backward scattering, Fig. \ref{fig:PhotonNeutronPos} shows four separate histograms representing the difference between photon and neutron CoM positions for events with 1, 2, 3, and 4 neutron scatters. A single layer thickness is 6~mm, represented by the magenta lines aligned at $\pm$3~mm in Fig. \ref{fig:PhotonNeutronPos}. Events in which the reconstructed photon and neutron CoM positions are within the same layer comprise 91\%, 86\%, 86\%, and 87\% of all events with 1, 2, 3, and 4 neutron scatters, respectively. 
\begin{figure}
  \centering
  \includegraphics[width=\columnwidth]{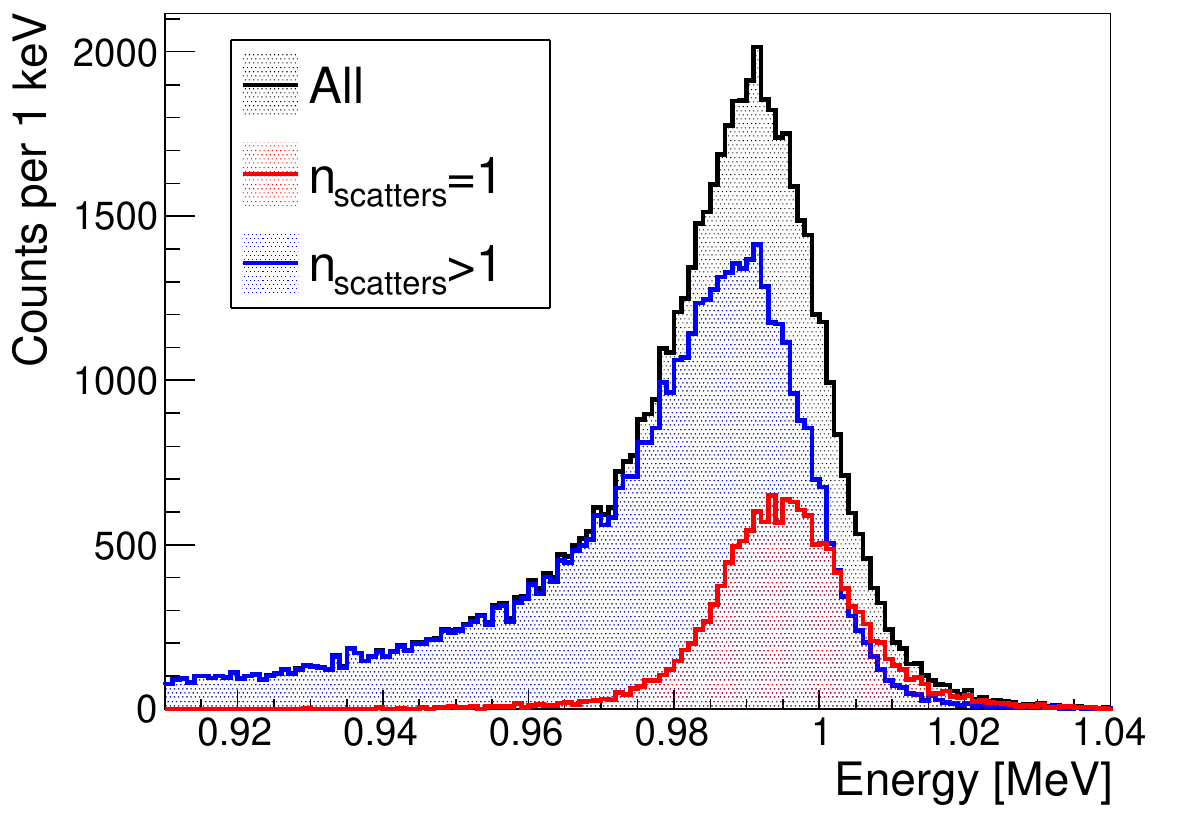}
  \caption{Calculated neutron energy distributions for 1~MeV neutrons over a 75~cm flight path. ToF was calculated using the weighted average photon arrival time for each photosensor. All simulated events are shown in the black histogram with relative contributions from single- and multiple-scattering events shown in the red and blue histograms, respectively.}
  \label{fig:simNeutronToF}
\end{figure}

To determine the overall effect of neutron multiple scattering on the detector performance, 1~MeV neutrons in a knife beam were simulated 75~cm away from a detector. The weighted average photon arrival time for each photosensor was used as a ToF measurement to calculate the incident neutron energies. In Fig. \ref{fig:simNeutronToF}, the black histogram represents the calculated neutron energies for all simulated 1~MeV neutrons, with relative contributions from single- (red) and multiple-scattering (blue) events. The maxima of the black, red, and blue distributions are 0.992, 0.994, and 0.990~MeV, respectively. The change in peak location between singly- and multiply-scattered neutrons shows the relationship between timing and reconstructed positions for multiple-scattering events is not as strongly correlated as for single-scattering events and typically results in lower reconstructed neutron energies. 

 Simulations of the NEXT prototype have shown that such detectors should be capable of measuring neutrons with improved energy resolution. The NEXT prototyping process was guided by single segment-simulations which minimized the effort needed to fully test every configuration with an experimental setup.
Going forward, the NEXT\emph{sim} framework will be continually developed to provide first estimates of new detector capabilities and simulate complete experiments with more sophisticated detector arrangements.

\section{Detector Prototyping}
\begin{figure}[bt]
\centering
\includegraphics[width=0.95\linewidth]{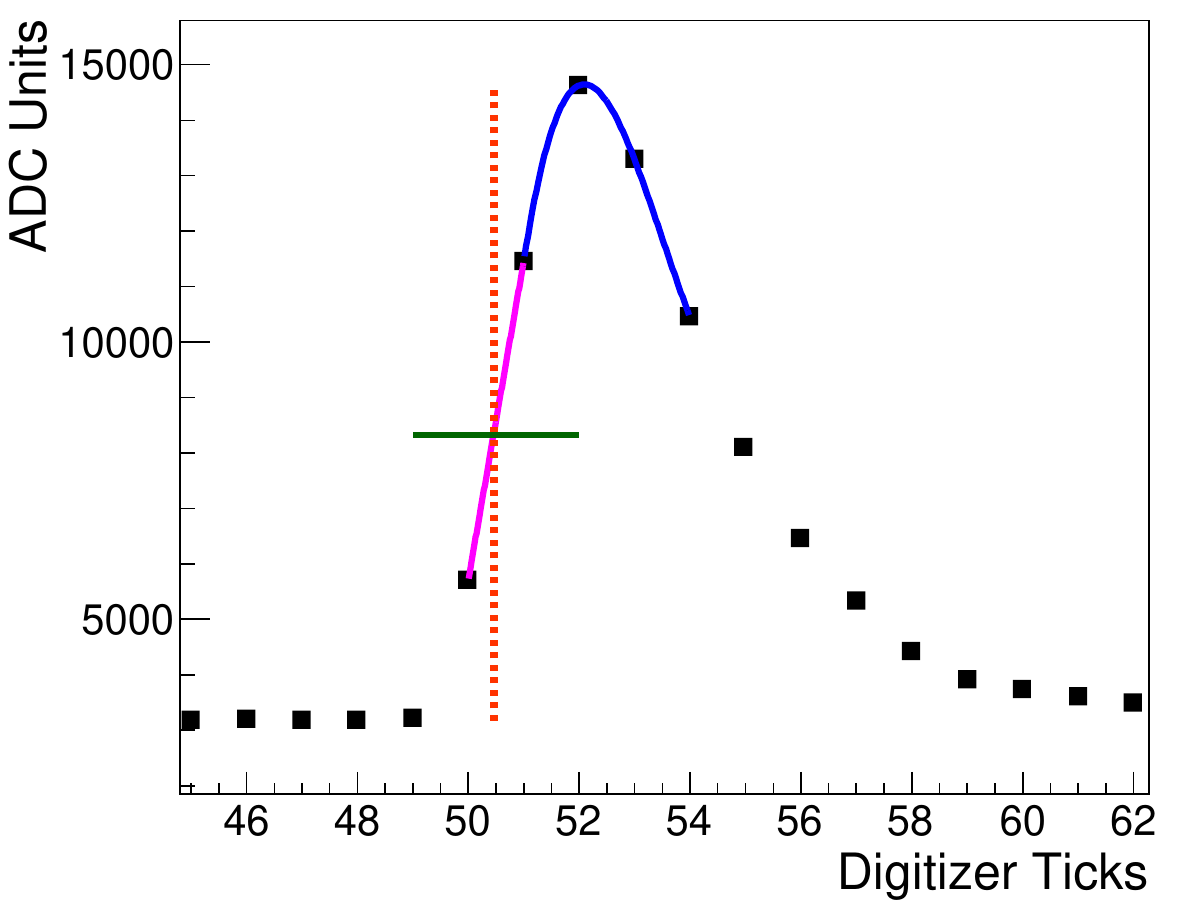}
\caption{Example of the PolyCFD algorithm applied to a digitized trace. The blue line represents the third-order polynomial fit to the maximum. The magenta line shows the linear interpolation, and the green line represents the CFD threshold level. The high-resolution time (HRT) is determined by the intersection between the magenta and green lines and is represented by the dashed red line.}
\label{fig:PolyCFD}
\end{figure}

The development phase of the NEXT project investigated single-segment scintillators of various geometries and different photosensors. The main goal of these tests was to explore whether the scintillation light produced by the interaction of neutrons in the plastic would be sufficient to retain the timing and PSD capabilities under particular design requirements.
\begin{figure*}[ptb]
  \centering
  \begin{subfigure}{\textwidth}
  \centering
  \includegraphics[width=0.9\textwidth,trim={0 5cm 0 2cm},clip]{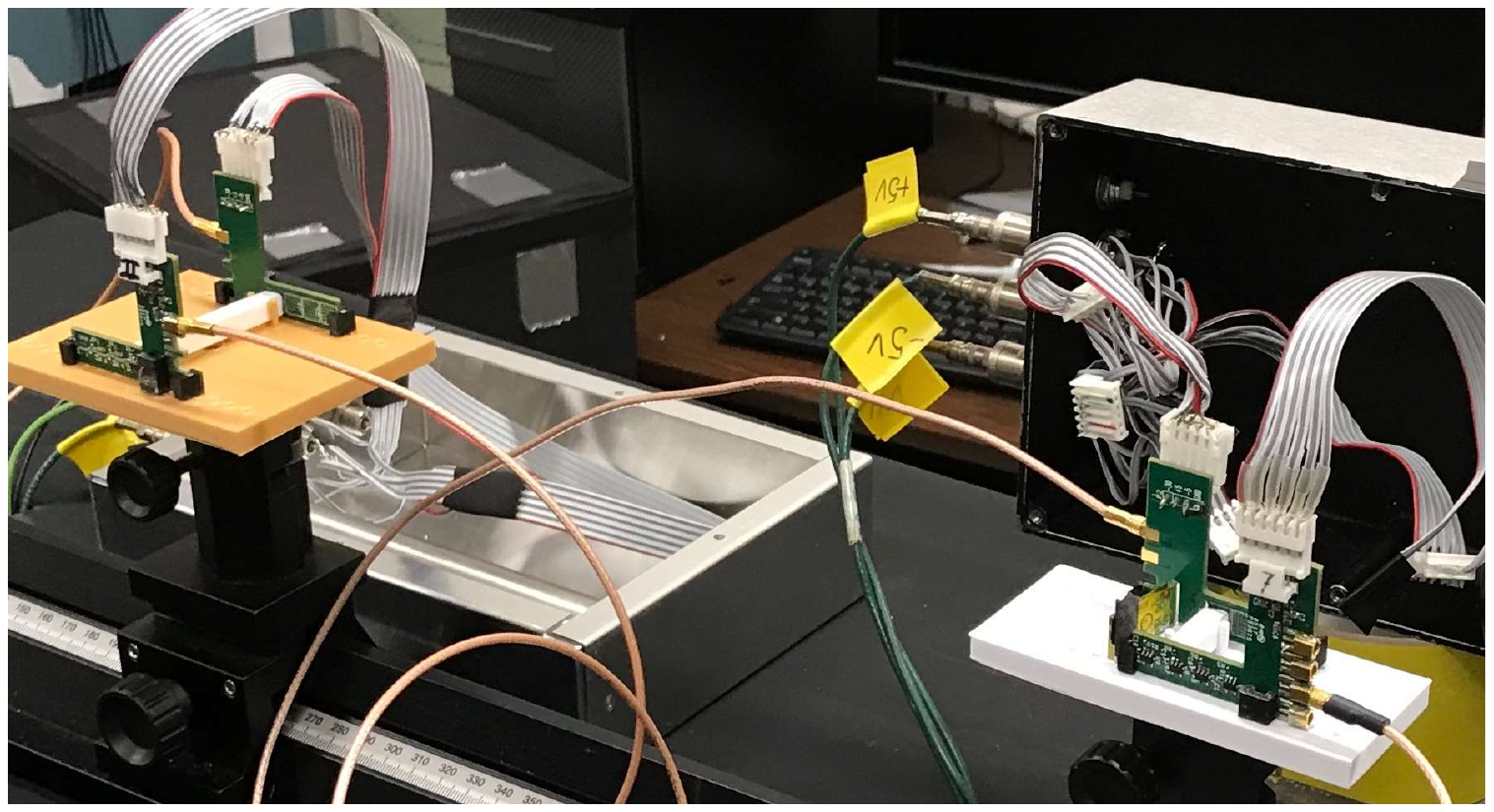}
  \caption{}
  \label{fig:SiPM_setup}
  \end{subfigure}%
  \\
  \begin{subfigure}{\textwidth}
   \centering
   \includegraphics[width=0.9\textwidth, trim={0 2cm 0 0}]{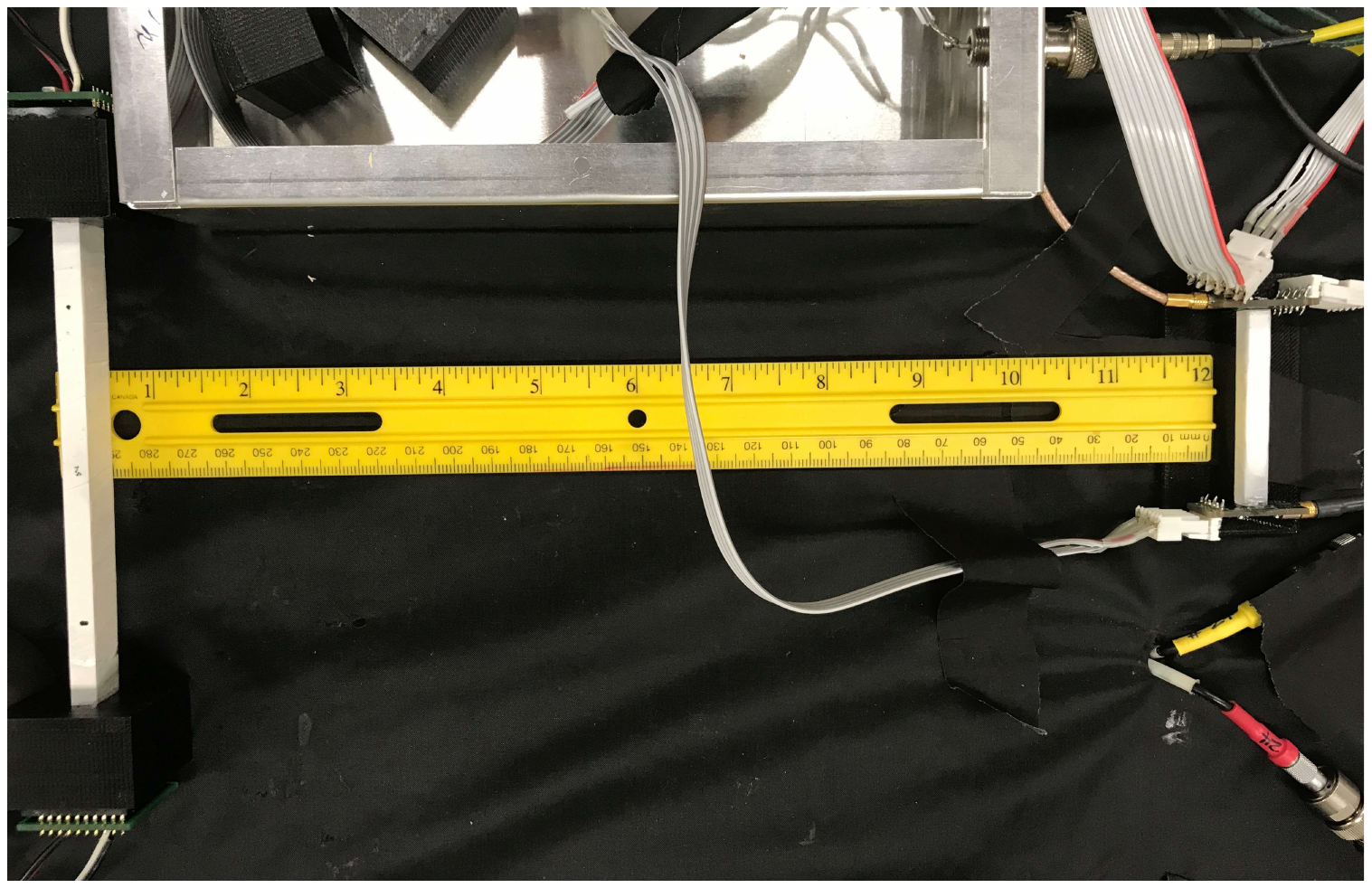}
   \caption{}
   \label{fig:EJ276_setup}
  \end{subfigure}%
  \caption{Two experimental setups for ToF tests using SiPM detectors (a) and EJ-276 scintillator (b). In each setup, a $^{252}$Cf source was placed next to the START detectors in the right side of the figures. The detectors to the left in the figures were the STOP detectors used to measure ToF for neutrons and gamma rays.}
  \label{fig:Setups}
\end{figure*}

\subsection{Detector and data acquisition}
Two main types of detector setups were tested to determine the applicability of different approaches for building a prototype. First, SiPM timing capabilities were studied using small scintillators attached to SiPMs. Second, bars of EJ-276 coupled to small, fast-timing PMTs were tested to determine the feasibility of incorporating PSD plastic.
The same data acquisition system (DAQ) was used to record and analyze signals for each experimental setup. The signal output from each photodetector was connected to 16-bit, 250~MHz (4~ns sampling) Pixie-16 digitizers developed by XIA LLC \cite{XIA} to digitize and store traces for later high-resolution timing analysis. A detailed description of a similar DAQ setup can be found in \cite{PAULAUSKAS201422}. The feasibility of each detector setup was determined by measuring left-right (axial) timing resolution and ToF resolution for both configurations.

\subsection{High resolution timing} \label{sec:HRT}
Neutron time-of-flight as well as the axial position of interaction along the scintillator can be determined using the time difference between the signals from photodetectors on either end of the detector. The internal timestamps of the Pixie-16 digitizers are in 8~ns intervals \cite{PixieManual} so a method to determine a more precise timing (sub-ns) was implemented.
The high resolution time (HRT) of each digitized pulse was determined by means of a polynomial constant fraction discrimination (PolyCFD) algorithm \cite{PhDCory}. The algorithm computes the maximum from a polynomial fit around the peak of the digitized pulse and the CFD threshold, $F$, is set as a fraction of the difference between the maximum and the baseline. The relative trace phase is equal to point where the linear interpolation between the points surrounding the CFD threshold in the leading edge crosses the threshold. The HRT for each trace is the sum of the trigger-latching timestamp and the relative trace phase, defined more clearly in Ref. \cite{PAULAUSKAS201422}. The optimal threshold fraction values were obtained for a factor range between $F=40-45\%$. A graphical representation of the PolyCFD method can be seen in Fig. \ref{fig:PolyCFD}.

\begin{figure}[tp]
  \centering
 \begin{subfigure}{\linewidth}
   \centering
   \hspace{-0.0cm}
  \includegraphics[width=0.95\linewidth]{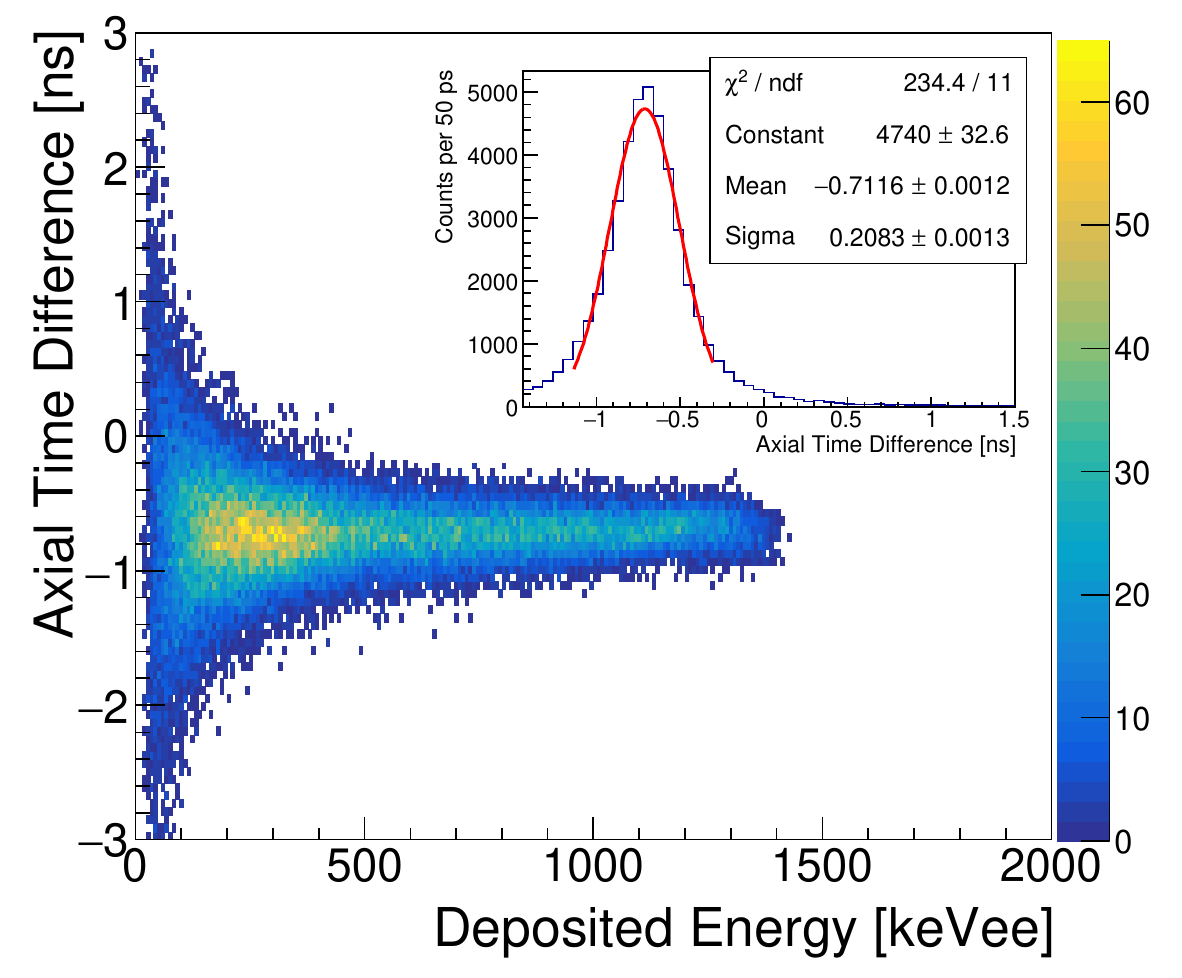}
  \caption{}
 \end{subfigure}%
 \\
 \begin{subfigure}{\linewidth}
  \centering
  \hspace{-0.0cm}
  \includegraphics[width=0.95\linewidth]{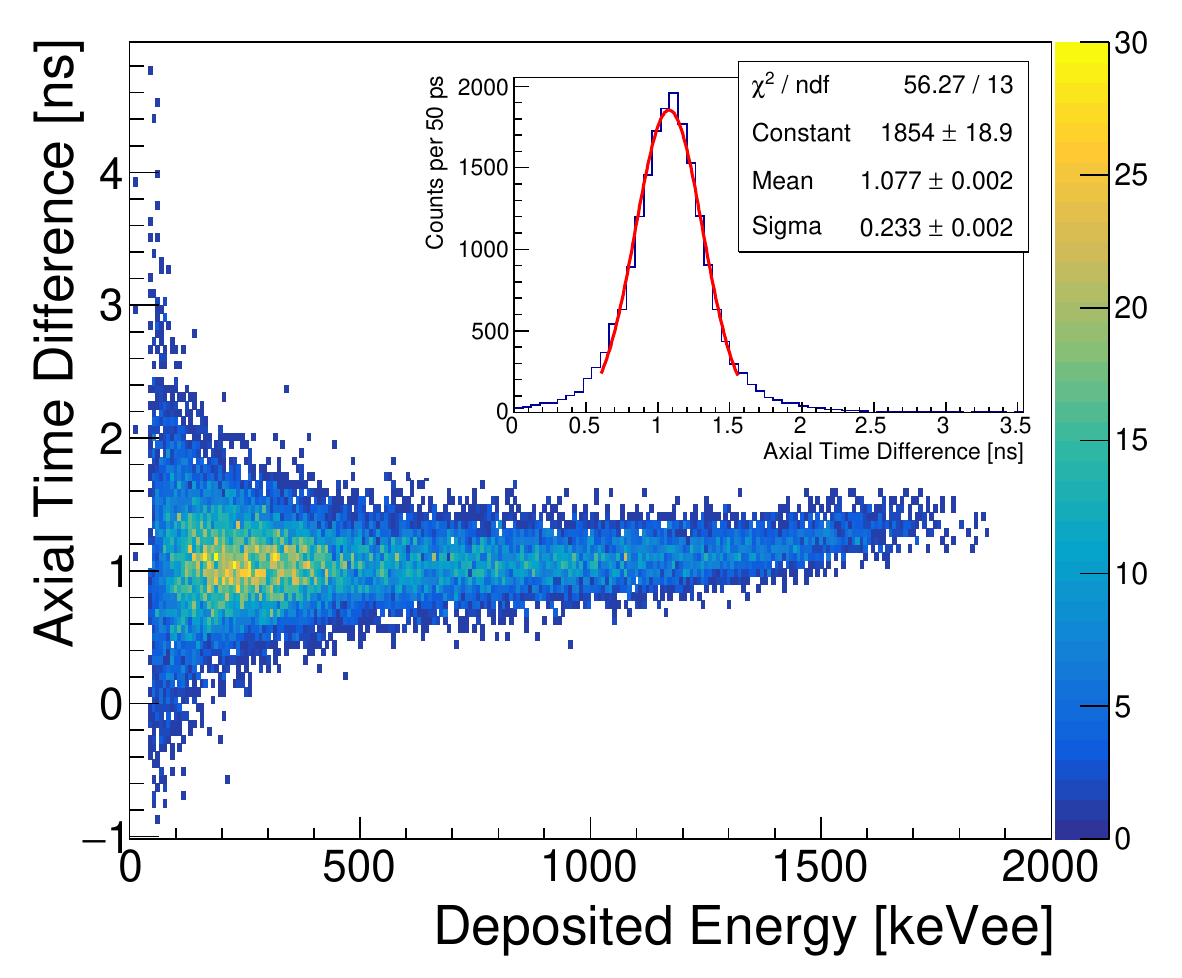}
  \caption{}
 \end{subfigure}%
 \caption{Two-dimensional histograms of the axial (left-right) time difference versus the deposited energy in a EJ-200 plastic scintillator from a $^{90}$Sr source, measured with on-board amplification (a) and without (b). The insets in each figure are time axis projections with Gaussian fits used to extract the FWHM timing resolution.}
 \label{fig:SiPMtiming}
\end{figure}

A $^{90}$Sr source was used to measure axial timing resolution by taking the difference of the left and right photosensor signal times. The $^{90}$Sr source provided a wide range of energy depositions up to $\sim$2~MeV. To determine ToF capabilities of the different detector setups, neutrons from a $^{252}$Cf fission source were measured and their respective flight times were measured between START and STOP detectors. The $^{252}$Cf source was placed as close as possible to the START detector to detect the gamma rays from de-excitation associated with the neutron emission. A neutron or another gamma ray was later observed in the STOP detector placed some distance away from the source. The $^{252}$Cf neutron spectrum is very well characterized \cite{Mannhart} and provides a good test of a detector's ToF capabilities. Images of the SiPM and EJ-276 setups for the ToF tests can be seen in Figs. \ref{fig:SiPM_setup} and \ref{fig:EJ276_setup}.

\section{Timing with SiPMs} \label{sec:SiPMs}

Silicon photomultipliers (SiPMs) offer a small form factor design and quantum efficiency uniformity for multi-detector arrays. SiPM left-right position and time-of-flight resolution was measured to ascertain the applicability of SiPMs to small-scale arrays. Two different SiPM readout circuits were designed to determine the effects of on-board filtering and amplification. The first one consisted of a simple low-pass active filter based on the Texas Instruments\textsuperscript{\textregistered} OPA656 operational amplifier recommended by SensL\textsuperscript{\textregistered}, the SiPM manufacturer \cite{JseriesUM}. A $25\Omega$ feedback resistor was chosen to maintain the fast rise-time of the SiPM signal while filtering high-frequency noise. The second circuit did not include on-board amplification. Pairs of identical SiPM signal readout boards were coupled to opposite ends of a PTFE-wrapped $50\times6\times6$~mm$^3$ piece of Eljen 200 (EJ-200) plastic scintillator. For each pair, the signals were gain matched and amplified using an ORTEC\textsuperscript{\textregistered} 535 fast amplifier module.

\subsection{SiPM axial timing resolution}
Fig. \ref{fig:SiPMtiming} shows the results of the axial timing (defined in Sect. \ref{sec:HRT}) measurements using the PTFE-wrapped EJ-200 and the $^{90}$Sr source with the two SiPM circuits mentioned previously. The top panel (a) corresponds to the circuit with active on-board amplification and the bottom panel (b) corresponds to the circuit without on-board amplification. The two-dimensional histograms in each figure show the left-right time difference plotted against the deposited energy in the plastic scintillator.
\begin{figure}[t]
  \includegraphics[width=\linewidth]{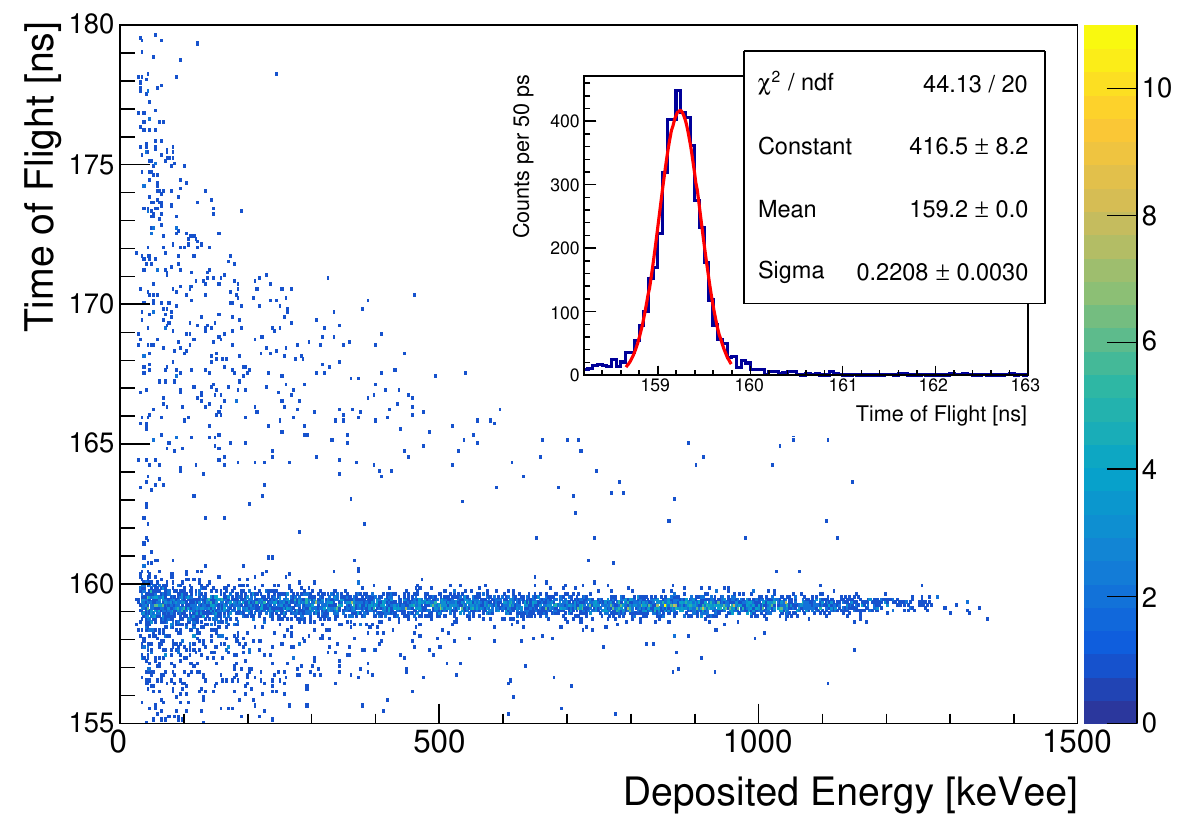}
  \caption{Two-dimensional histogram of $^{252}$Cf time-of-flight measurements plotted against deposited energy in the EJ-200 stop detector. The inset shows a projection of the gamma-ray peak onto the ToF axis. ToF resolution, $\Delta ToF$, was determined to be 518~ps from a Gaussian fit to the gamma peak. The ToF data are shown here with no offset to account for inherent timestamp differences between START and STOP acquisition channels.}
  \label{fig:ToF_SiPM}
\end{figure}
The insets in each panel of Fig. \ref{fig:SiPMtiming} show projections on the time axis of the histograms. Timing resolution, $\Delta t$, is calculated as the full width-half maximum (FWMH) of Gaussian fits to time difference distributions ($\Delta t=2.35\times\sigma_{t}$). The left-right timing resolutions obtained for the filtered/amplified board and the non-amplified board are $\Delta t=489$~ps and $\Delta t=548$~ps, respectively. Once the axial-timing capabilities of a SiPM-based detector were validated, a small ToF setup was made to test SiPM's applicability to such a configuration. 

\subsection{Time-of-flight with SiPMs} \label{sec:SiPMTOF}
As a proof-of-principle, a small scale ToF configuration was implemented to measure neutron flight times from a $^{252}$Cf source. The setup consisted of a $20\times6\times6$~mm$^3$ piece of EJ-200 plastic scintillator attached to $6\times6$~mm$^2$ SensL\textsuperscript{\textregistered} SiPMs used as a START detector and a $50\times6\times6$~mm$^3$ EJ-200 bar attached to identical SensL\textsuperscript{\textregistered} SiPMs used as the STOP detector, placed 29~cm apart. An image of SiPM ToF setup can be seen in Fig. \ref{fig:SiPM_setup}. Both bars of EJ-200 were wrapped in PTFE tape. The same acquisition and timing methods from the axial timing resolution tests were used. A ToF vs. deposited energy distribution can be seen in Fig. \ref{fig:ToF_SiPM}, the inset showing a 1-D projection on the ToF axis. The ToF data shown in Fig. \ref{fig:ToF_SiPM} have not been corrected for inherent timestamp differences between the START and STOP detectors due to the configuration of the specialized triggering firmware. Time-of-flight resolution is defined similar to axial timing resolution, $\Delta ToF = 2.35\times\sigma_{ToF}$. From a Gaussian fit to the gamma-ray peak in the Fig. \ref{fig:ToF_SiPM} inset, $\Delta ToF=518$~ps for the SiPM setup.
\begin{figure}[bt]
  \centering
 \includegraphics[width=\linewidth]{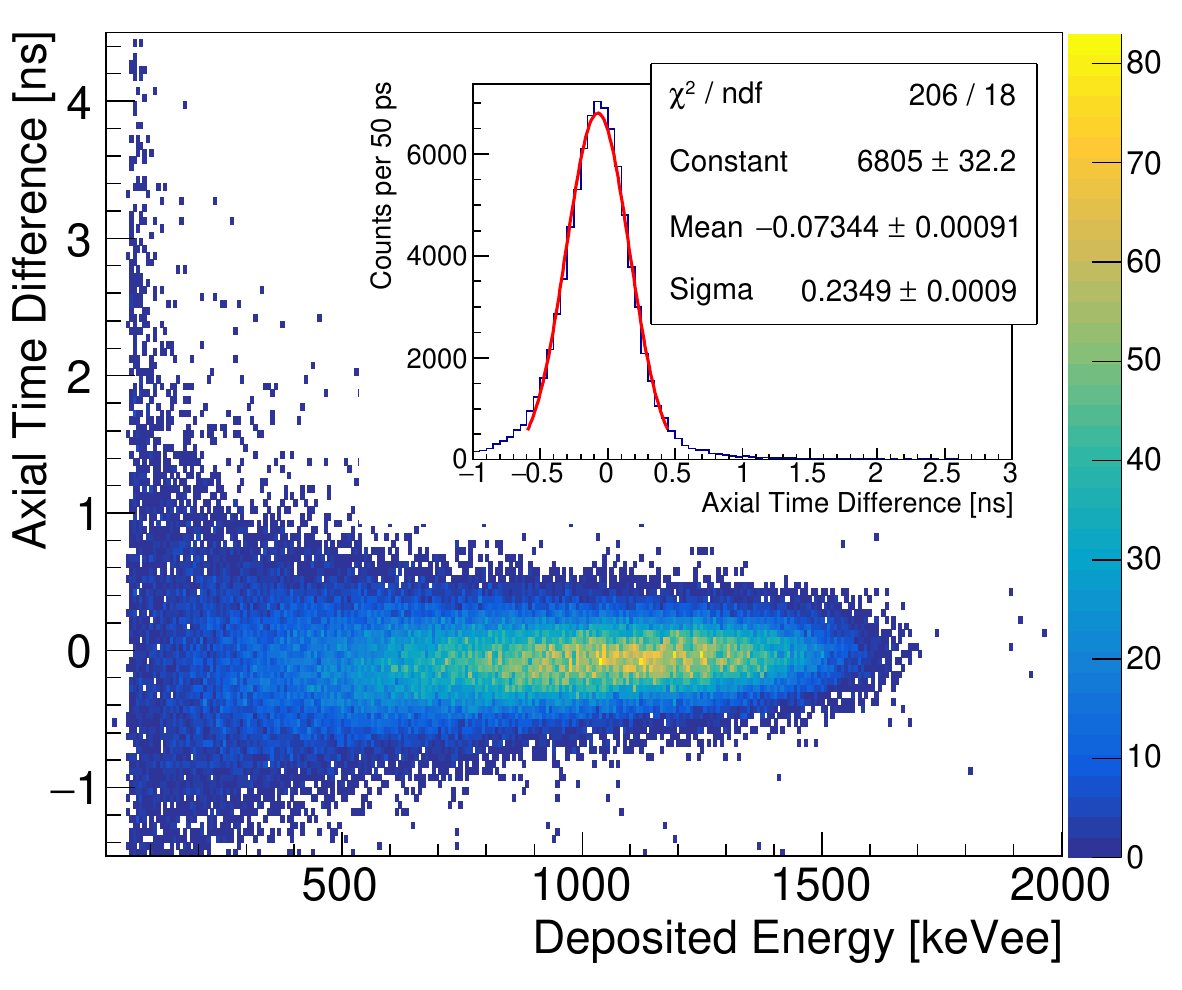}
  \caption{Two-dimensional histogram of the time difference between PMTs on opposite ends of a 254~mm ESR-covered bar of EJ-276. Inset: Projection of the two-dimensional histogram on the time axis. A Gaussian fit (red) to the distribution yields axial timing resolution $\Delta t$=552~ps.}
  \label{fig:MylarTiming}
\end{figure}

The small-scale SiPM timing tests establish SiPMs as viable detectors for small-scale ToF arrays. Testing will continue to determine the scalability of SiPMs to a large, multi-detector resistive-readout system.

\section{Eljen 276 Detector Tests} \label{sec:EJ276}

\subsection{Timing tests with PMTs}
\begin{figure}[tp]
  \centering
 \includegraphics[width=\linewidth, trim = {0 0 0 0}, clip]{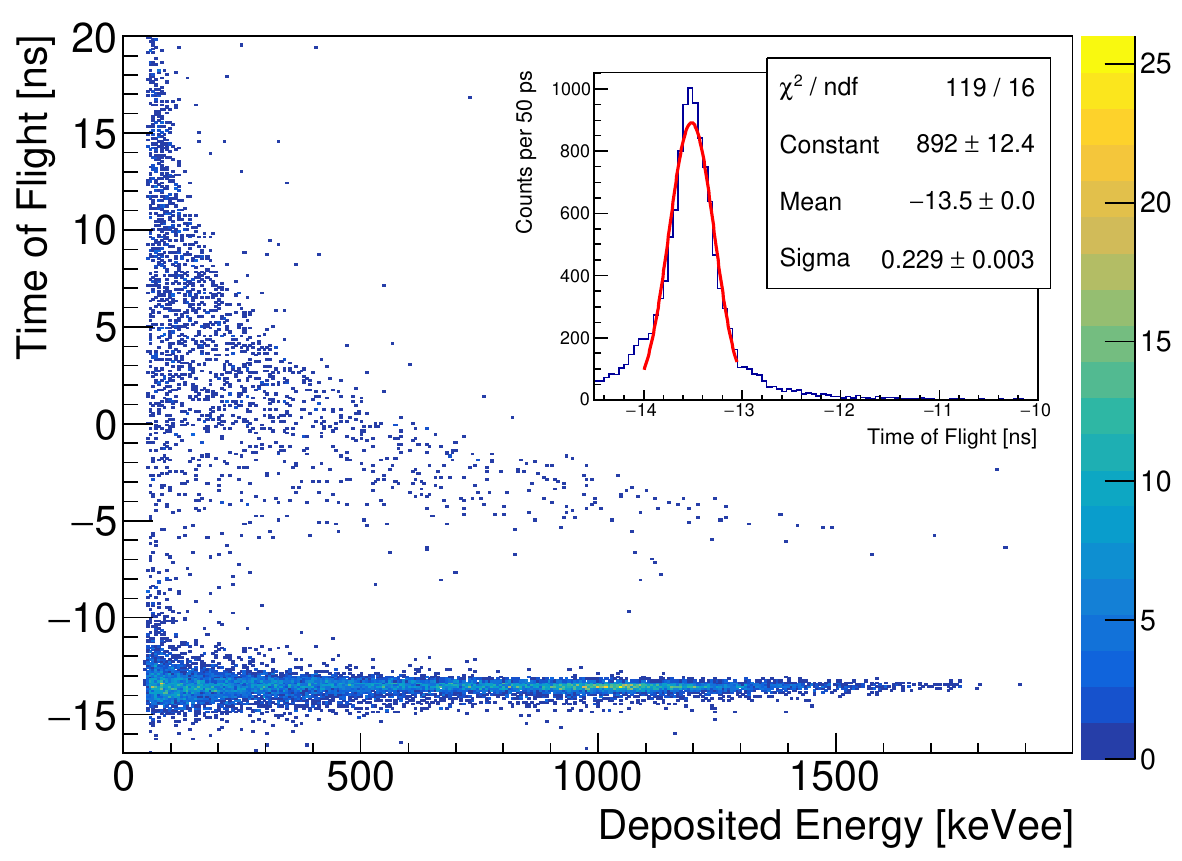}
 \caption{Two-dimensional histogram of $^{252}$Cf ToF versus deposited energy in the ESR wrapped EJ-276 stop detector. The inset is a projection of the gamma-ray peak in the ToF spectrum and has $\Delta ToF$=538~ps (50~keVee threshold). The ToF data are shown here with no offset to account for inherent timestamp differences between START and STOP acquisition channels.}
 \label{fig:TOFEJ276}
\end{figure}

 Timing performance of $127\times12.7\times6$~mm\textsuperscript{3} EJ-276 bars was measured using fast, compact Hamamatsu R11265U PMTs \cite{HMPMT}. The 6~mm thick bars of EJ-276 were machined from 0.5$\times$11$\times$12~in$^3$ sheets by Agile Technologies, Inc. \cite{AGILE} and the sides of each bar were covered with either 3M\textsuperscript{\texttrademark} ESR (Enhanced Specular Reflector) \cite{3MESR} or Lumirror\textsuperscript{\texttrademark} (produced by Toray) \cite{TORAY}. ESR is a specular reflector with 98\% reflectivity in the visible spectrum and Lumirror\textsuperscript{\texttrademark} is a diffuse reflector (reflective properties similar to PTFE). Both wrappings were applied to the EJ-276 bars using a UV-cured optical adhesive. The pre-covered bars were also wrapped in PTFE-tape to prevent light leakage at the edges. Bars were also provided with no wrapping to check the effect of the reflective layers. 

\begin{figure*}[tbp]
 \centering
  \includegraphics[width=0.98\textwidth]{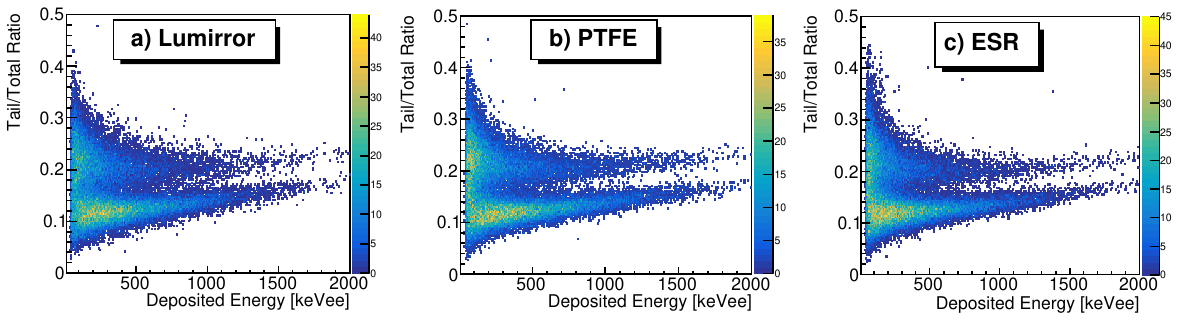}
  \caption{Two-dimensional histograms of the CCM PSD for three different types of wrapping.}
  \label{fig:PSDEJ276}
\end{figure*}

Fig. \ref{fig:MylarTiming} shows the two-dimensional histogram of the axial time difference plotted against the deposited energy in the ESR-wrapped EJ-276 scintillator. The inset plot in Fig. \ref{fig:MylarTiming} shows the y-axis projection of the axial time difference fit with a Gaussian distribution, yielding $\Delta t=543$~ps. The Lumirror\textsuperscript{\texttrademark} wrapped detector was not tested for timing due to the poor n-$\gamma$ discrimination that is detailed in Sect. \ref{sec:PSD}.
A ToF setup similar to the SiPM test (Sect. \ref{sec:SiPMTOF}) was made to measure the ToF resolution for a single 127~mm long bar of EJ-276 plastic scintillator wrapped with ESR. An image of the ToF setup can be seen in Fig. \ref{fig:EJ276_setup}. The $^{252}$Cf ToF spectrum was measured using a $50\times6\times6$~mm$^3$ PTFE-wrapped EJ-200 bar attached to amplified SiPMs as the START detector and the EJ-276 bar coupled to PMTs as the STOP detector (oriented with the 6~mm thickness along the direction of the incident neutrons). The ToF vs. deposited energy results are shown in Fig. \ref{fig:TOFEJ276}. The inset plot shows a Gaussian fit to the gamma-ray peak. As in Sect. \ref{sec:SiPMTOF}, the ToF data shown in Fig. \ref{fig:TOFEJ276} have not been corrected for inherent timestamp differences between the START and STOP detectors. The resolution was determined to be $\Delta ToF=538$~ps from the Gaussian fit when a 50~keVee threshold was applied.

\subsection{Neutron-gamma discrimination} \label{sec:PSD}
EJ-276 evolved from first-generation plastic scintillator with n-$\gamma$ discrimination capabilities, EJ-299. The n-$\gamma$ response mechanism for EJ-276 is accurately described in Ref. \cite{ZAITSEVA201897}. 
\begin{figure}[t]
  \centering
 \begin{subfigure}{\columnwidth}
  \centering
	 \includegraphics[width=0.95\linewidth,trim={0cm 0 0cm 0},clip]{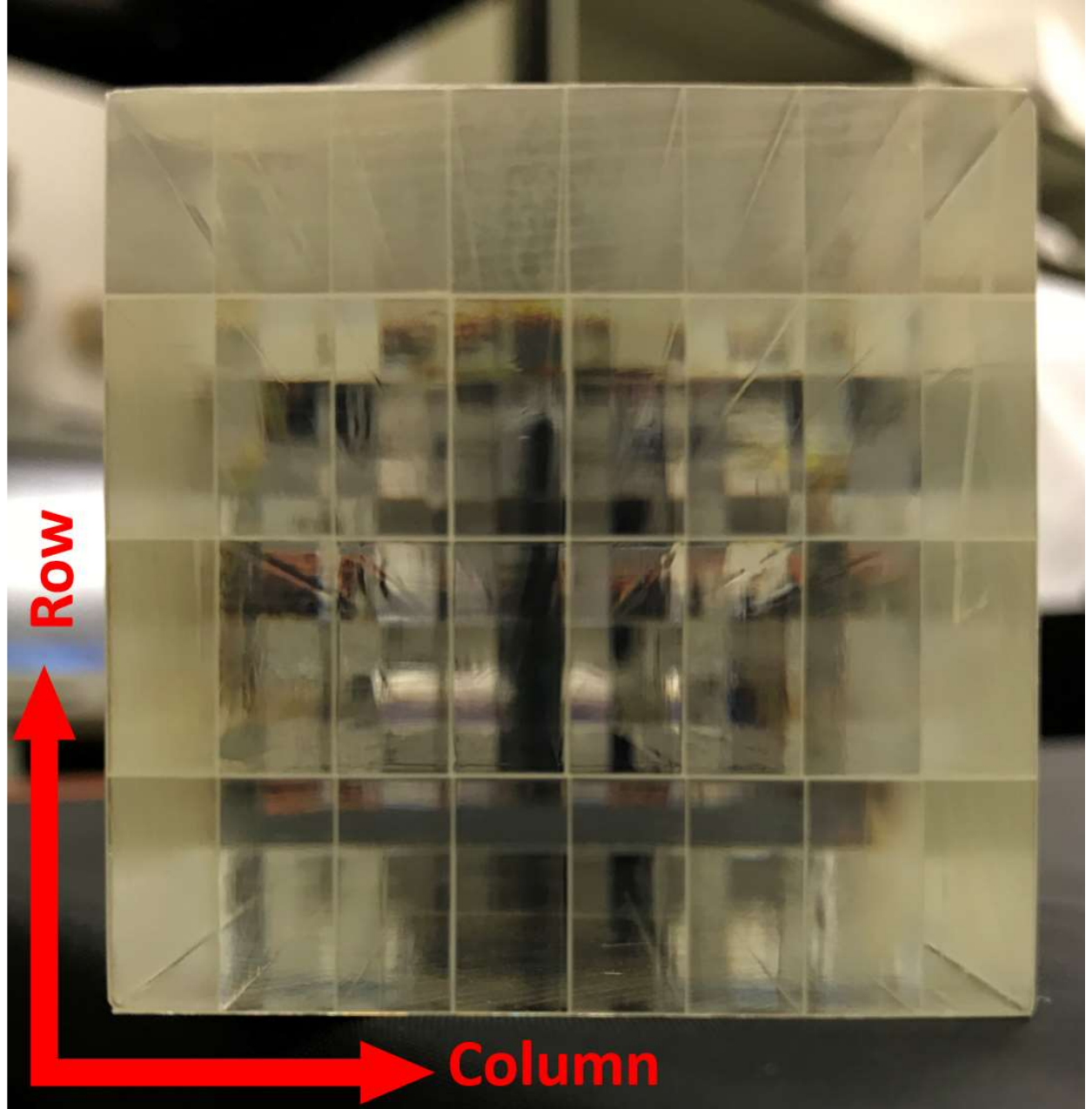}
 \end{subfigure}%
  \\
\begin{subfigure}{\columnwidth}
  \centering
  \includegraphics[width=0.95\linewidth]{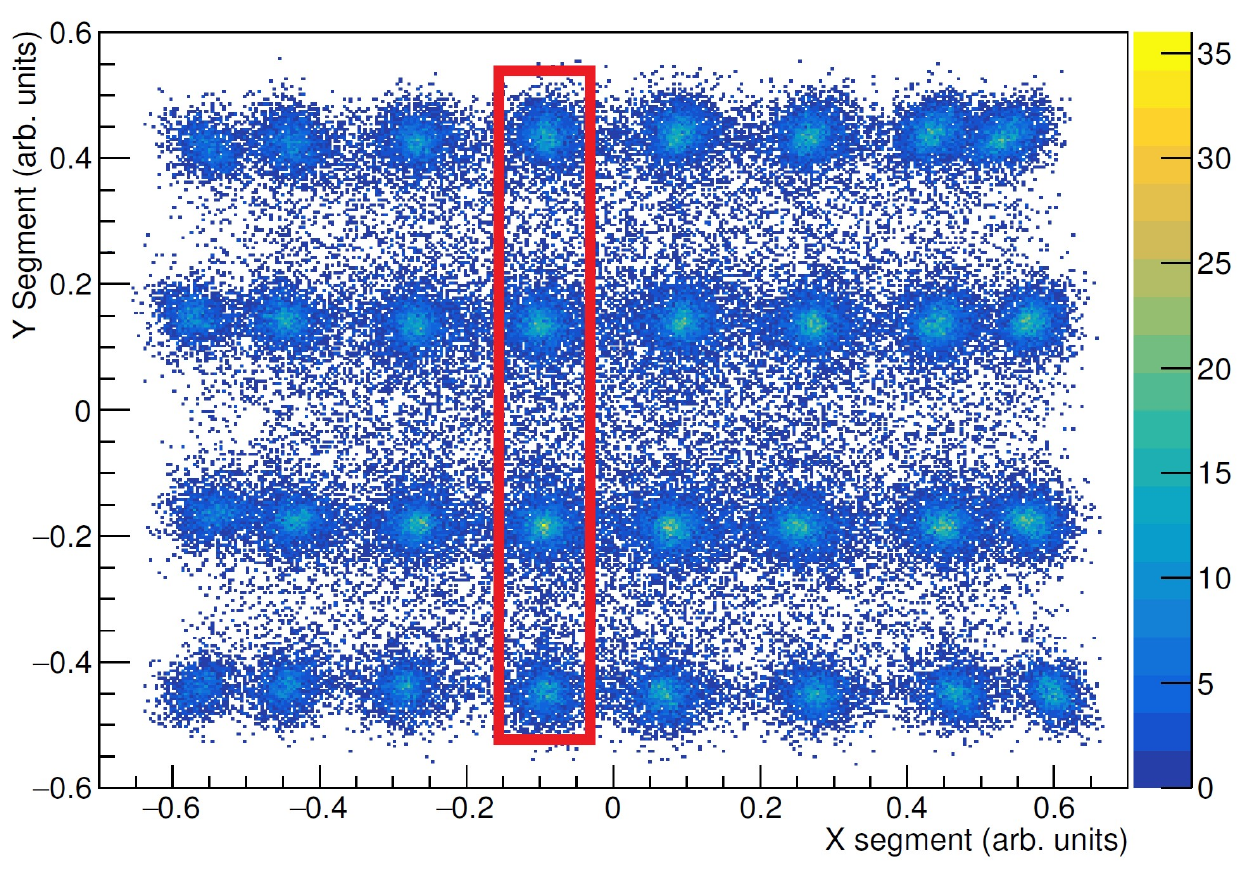}
\end{subfigure}%
\caption{The top figure is an image of one end of a 4$\times$8 segmented scintillator. The bottom figure shows the reconstructed cells using the position sensitive signals from the Vertilon interface board. A single column is outlined by the red rectangle in the bottom figure. The detector is always arranged such that the higher segmentation is along the incident particle flight path for the best position and timing resolution.}
  \label{fig:PSPMTImage}
\end{figure}
The two pre-covered EJ-276 bars from Agile Technologies, Inc. (ESR and Lumirror\textsuperscript{\texttrademark}) and a third EJ-276 bar wrapped with PTFE were tested to measure the effect of different surface reflectors on the quality of the PSD. PTFE-wrapped bars were included only to determine the effect of the UV-cured optical adhesive; This wrapping can't be used in the construction of a multi-segmented module. Differences between ESR- and Lumirror\textsuperscript{\texttrademark}-covered bars will determine how specular and diffuse reflectors maintaining n-$\gamma$ discrimination information. EJ-276 scintillation light from scattered neutrons and gamma rays was recorded with the same Pixie-16 digitizer used in the earlier setups, and the PSD was tested using the charge comparison method (CCM) \cite{CCMPSD}. By measuring the total and partial (tail) integral of each signal and calculating the ratio between the two integrals, neutron events can be separated from gamma-rays events. This approach is further improved by calculating the geometric mean of the ratios from the left and right detector: $$r_{total}=\sqrt{r_{L} \times r_{R}}.$$ CCM has previously been determined to be optimal when using a high-bit-resolution digitizer \cite{HighResPSD}.

Using the $^{252}$Cf source and a 5~cm block of lead to attenuate the large gamma-ray flux, the waveforms from each bar were digitized and tail to total integral ratios were calculated. Fig. \ref{fig:PSDEJ276} shows the PSD plots for bars covered with Lumirror\textsuperscript{\texttrademark} (a), PTFE (b), and ESR (c) with figures of merit (FoMs) \cite{CCMPSD} being calculated between 400 and 500~keVee. The FoMs for the Lumirror\textsuperscript{\texttrademark}, PTFE, and ESR bars were 0.820$\pm$0.012, 1.042$\pm$0.016, and 0.977$\pm$0.015, respectively. The difference in FoM between the PTFE-wrapped and Lumirror\textsuperscript{\texttrademark} covered bars shows the optical adhesive worsens PSD capabilities. Of the two possible prototype reflective coverings, ESR was better at maintaining PSD so further tests of EJ-276 segments were only done with ESR wrapping.

\section{NEXT Prototype} \label{Prototype}
ESR-wrapped EJ-276 segments were shown to meet NEXT design goals, leading to the assembly of 48$\times$50.8$\times$254~mm$^{3}$ segmented detectors. An individual segment or cell is 6$\times$12.7$\times$254~mm$^3$. A whole detector has 4$\times$8 scintillator cells, shown in Fig. \ref{fig:PSPMTImage}, the higher segmentation being along the direction of incident particles. A full NEXT prototype is made up of one 4$\times$8 segmented scintillator coupled to Hamamatsu H12700A position sensitive PMTs (PSPMTs) on each end of the segmented scintillator. The H12700A PSPMTs have an 8$\times$8 segmentation (6$\times$6~mm$^{2}$ anodes), each anode having an individual readout. A Vertilon PSPMT Anger Logic interface board (Model SIB064B-1018) \cite{VERTILON} was used to reduce the position sensitive readout from 64 individual position signals to 4 position signals, one at each corner of the SIB064B-1018 resistive network. This substantially reduces the total number of required DAQ channels.

The scintillation position is reconstructed using the weighted average of the 4 corner resistive network signals based on their respective integrated signals \cite{ANGER}. Fig. \ref{fig:PSPMTImage} shows the reconstructed scintillator segmentation using the Anger Logic position measurement of the PSPMTs from a measurement using a $^{60}$Co source. The PSPMT common dynode signal is connected directly to the acquisition and used for all timing and PSD analyses. The scintillator cell-dependent analysis calculates neutron energies on a segment-by-segment basis using reconstructed positions for the particle flight path.

\subsection{Time-of-flight measurements}
\begin{figure}[tp]
 \centering
 \includegraphics[width=0.99\linewidth,trim={0 0 0 0},clip]{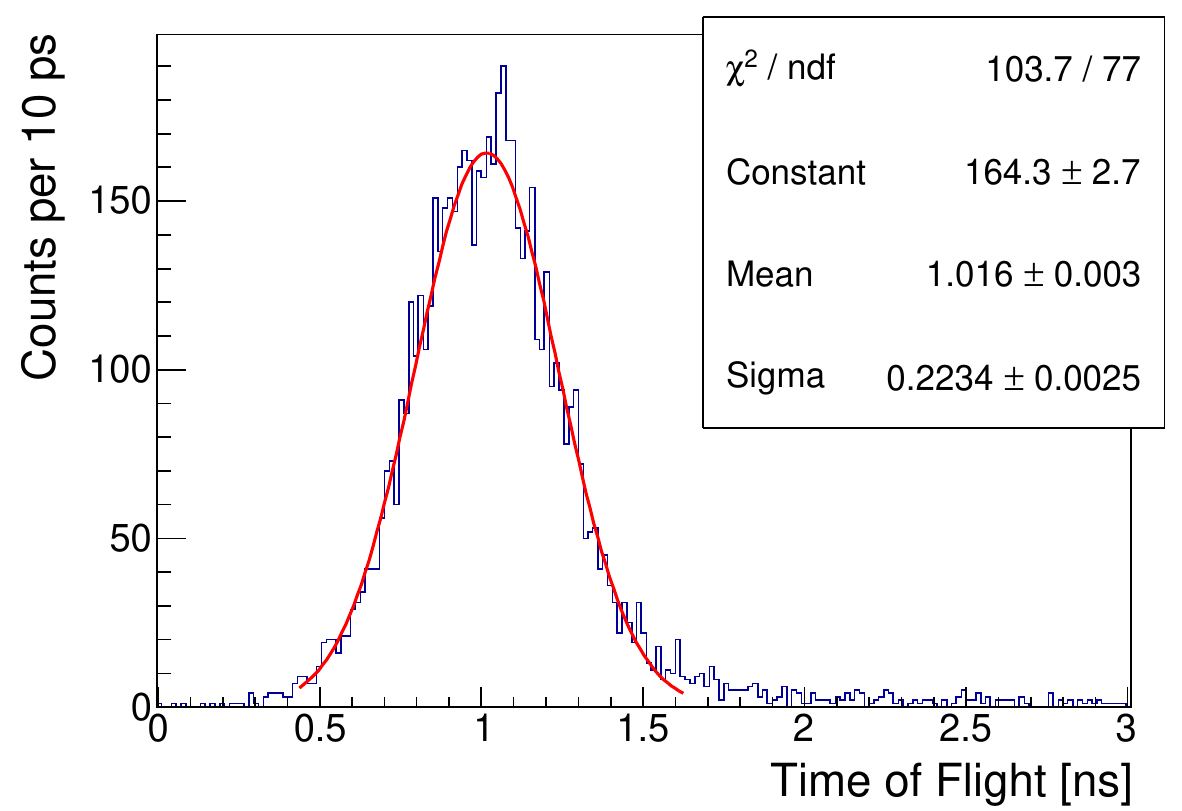}
 \caption{Time-of-flight resolution for a single NEXT prototype column (outlined in red in Fig. \ref{fig:PSPMTImage}) using a collimated $^{60}$Co source at a flight distance of $\sim$44~cm. The Gaussian fit to the distribution shows the time resolution is 525~ps (30~keVee threshold).}
 \label{fig:CollimatedCoToF}
\end{figure}
\begin{figure}[tb]
  \centering
  \includegraphics[width=0.9\linewidth]{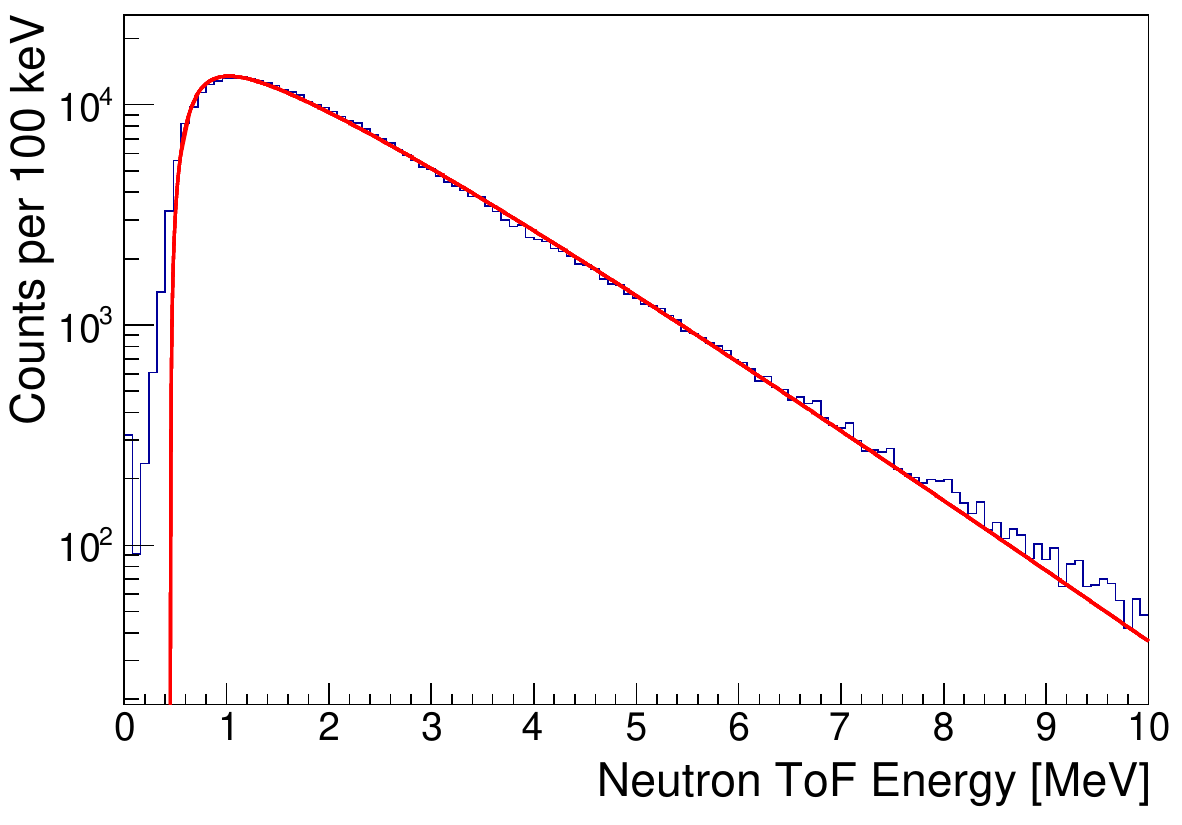}
  \caption{$^{252}$Cf neutron energy spectrum as measured with the NEXT prototype using the segment-dependent analysis (blue). The red line shows the expected neutron yield with a 100~keVee threshold.}
  \label{fig:Cf252Spectrum}
\end{figure}

\begin{figure}[t]
 \centering
 \includegraphics[width=0.99\linewidth]{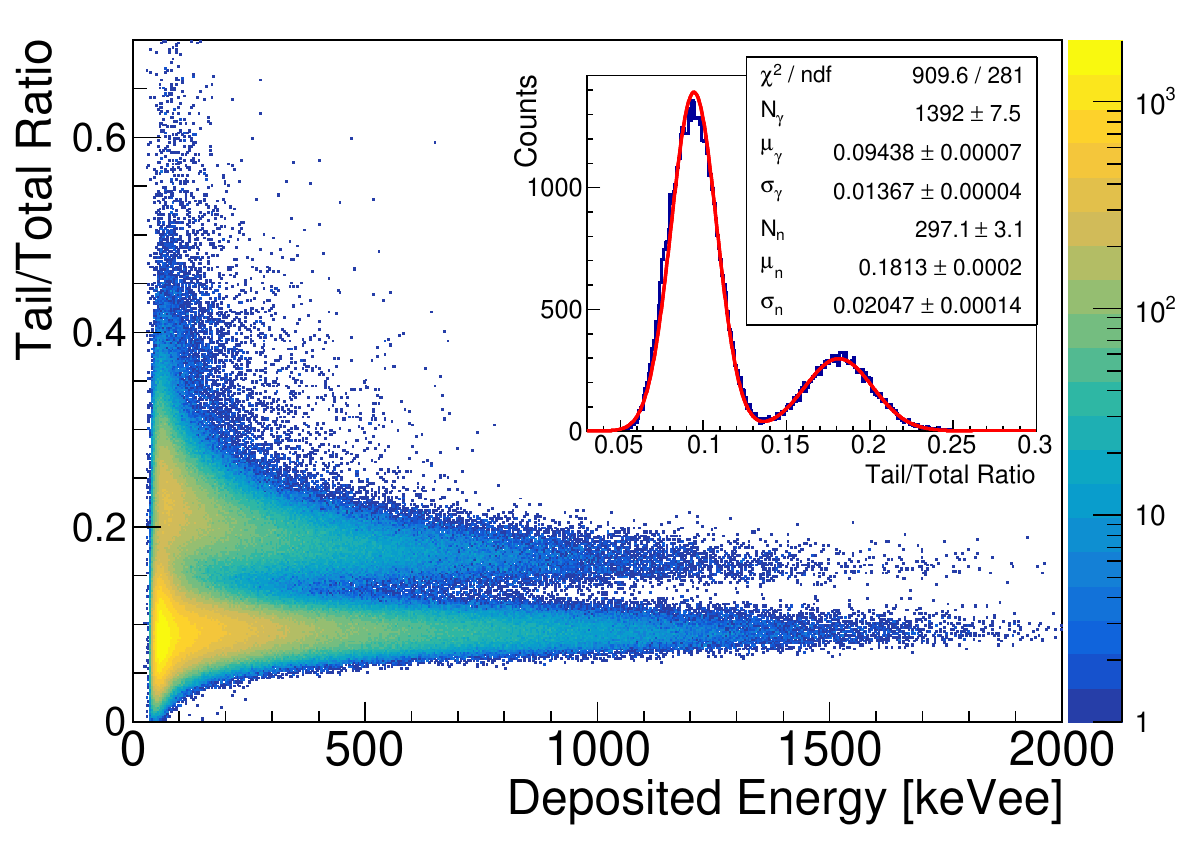}
 \caption{PSD from the common dynode signal of the PSPMT. The distribution has been corrected by adjusting the gamma-ray portion to have no linear dependence on deposited energy. The inset shows the PSD projection in the deposited energy range of 400-500~keVee. The FoM for this energy range is $1.08\pm0.02$.}
 \label{fig:PSPMTPSD}
\end{figure}

To measure the prototype time-of-flight resolution for a single column (as outlined in red in the bottom of Fig. \ref{fig:PSPMTImage}), a collimated $^{60}$Co source was used. Fig. \ref{fig:CollimatedCoToF} shows the ToF distribution for a single 6~mm column. From a Gaussian fit to the distribution, the ToF resolution is $\Delta ToF$=543~ps after applying a 30~keVee threshold. Once the prototype was established to have $\sim$500~ps ToF resolution for a single layer, a proof-of-principle neutron energy measurement was made using a $^{252}$Cf source. The source was placed $\sim$44~cm from the front face of the prototype. The neutron yield, shown in Fig. \ref{fig:Cf252Spectrum}, was calculated using time-of-flight information. Using the PSD information shown in Fig. \ref{fig:PSPMTPSD}, a cut was applied to select only neutron events. The data were fit with the Watt equation (red line in Fig. \ref{fig:Cf252Spectrum}) using fixed Mannhart parameters \cite{Mannhart}. An additional multiplicative factor was also used to scale the fit to the data. The disagreement at low neutron energy with respect to (w.r.t.) the fit is likely due to a stringent detection threshold in simulated efficiency data which was folded with the Watt equation. 

\subsection{Neutron-gamma discrimination}
The PSPMT single photon response is different than that of the fast timing PMTs used to initially test EJ-276 and is not uniform across all anodes in PSPMT. This response affects the overall pulse shape, potentially affecting PSD capabilities. PSD can only be calculated using the dynode signals as the four position signals lose n-$\gamma$ information after passing through the resistive network. Fig. \ref{fig:PSPMTPSD} displays the prototype PSD capabilities using the CCM, the inset showing the PSD projection for the same energy cut (400-500~keVee) used in the single bar wrapping tests. In this energy window, the FoM is 1.08$\pm$0.02. The NEXT protype does not show any noticeable effect on PSD due to segmentation or multi-anode readout.

\section{Monoenergetic Neutron Tests}
NEXT's defining characteristic is high-precision, position-dependent timing correlations. When neutrons pass through the segmented detector, there is a non-negligible amount of time taken to traverse the thickness of a single column. For monoenergetic neutrons, ToF measurements should, therefore, correspond to the position within the detector the neutron interacted, i.e., the average ToF for each successive column should shift by the time it takes a neutron to traverse a single column thickness. In order to benchmark the timing-position correlation for the NEXT prototype, monoenergetic neutrons were measured at the University of Kentucky Accelerator Laboratory (UKAL).

\subsection{Experimental Setup at UKAL}
At UKAL, monoenergetic neutrons are generated with $^3$H($p,n$)$^3$He, $^2$H($d,n$)$^3$He, or $^3$H($d,n$)$^4$He reactions. An in-depth overview which describes the neutron production and energy selection at UKAL can be found in Ref. \cite{HARTMAN2015137}. The $^3$H($p,n$)$^3$He reaction was used to generate neutrons with energies in the 0.25 to 1.5~MeV range, but only $\sim$1~MeV neutrons will be discussed below as an example of NEXT's position-dependent timing characteristics. Subsequent publications will discuss observed ToF distribution shifts at other energies and any energy-dependent position timing correlations.

NEXT was positioned behind stacked copper, polyethylene/lead, and paraffin/lithium carbonate collimators and aligned at 55\textdegree~w.r.t. the proton beam direction, corresponding to 1.03(3)~MeV neutrons. 
Neutron ToFs were measured as the HRT difference between the proton beam pickoff signal, upstream of the tritium target, and the dynode timing signals from the NEXT prototype. Specialized XIA LLC Pixie-16 firmware allowed the acquisition to be run in triple coincidence mode \cite{VandleManual}, requiring a start signal (proton beam pickoff) and two stop signals (left-right dynode timing signals) within a pre-determined coincidence window.
\begin{figure}[t]
  \includegraphics[width=\linewidth, trim={0 0 0 0}, clip]{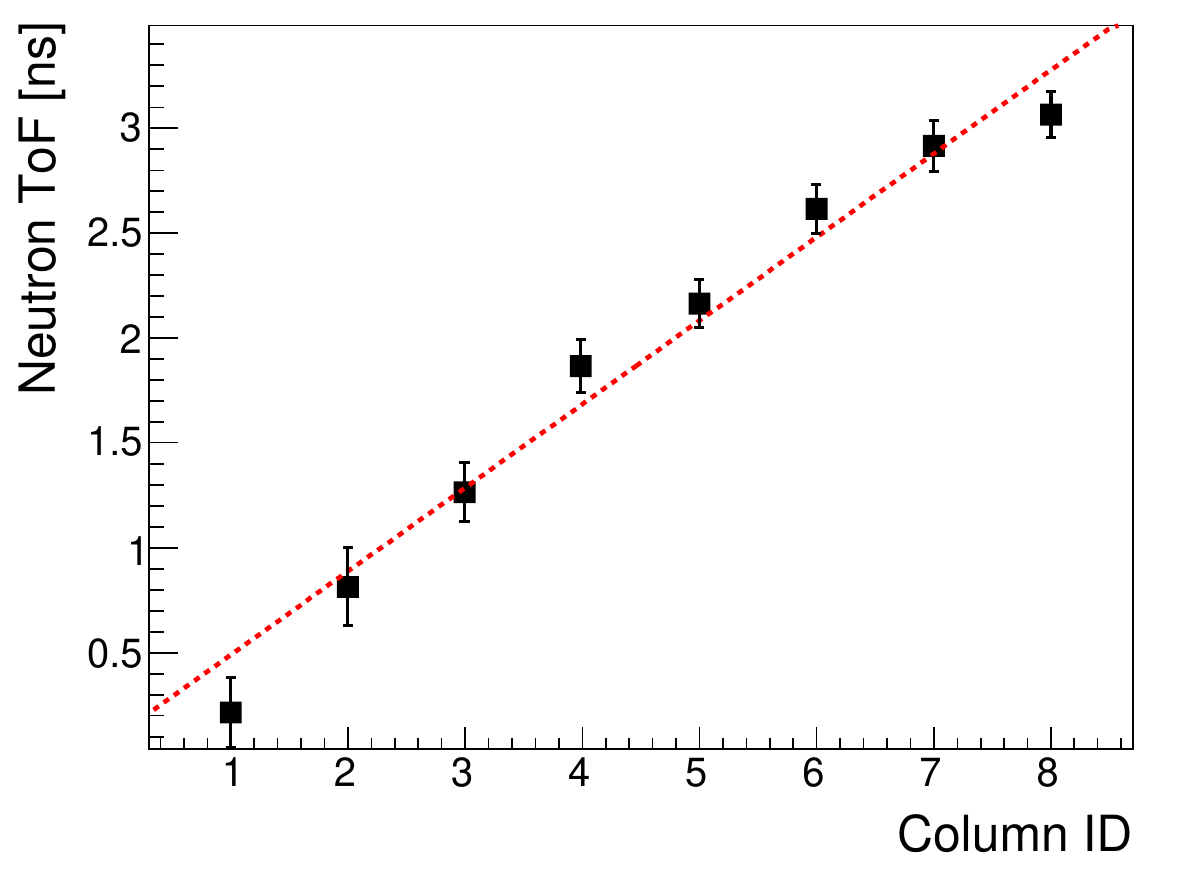}
  \caption{Mean ToFs for each segment from simulated 1.03~MeV neutrons, mimicking the UKAL experimental setup using the NEXT\emph{sim} framework. The dashed red line shows the expected position dependence of the ToF measurements for 1.03~MeV neutrons. The error associated with each point is the statistical uncertainty in the mean of the Gaussian fit to the ToF distribution for each segment.}
  \label{fig:simToFvsSeg}
\end{figure}

\subsection{Simulating ToF propagation}
 Ideally, the shift in the mean of ToF distributions for each successive segment would be constant. A simulation replicating UKAL NEXT measurements was completed to provide an estimate of the detector response to $\sim$1~MeV neutrons. 1.03~MeV neutrons in a cylinder beam with radius 25.4~mm were simulated along a 3.08~m flight path to the front of a NEXT prototype. Only 1.03~MeV neutrons were simulated because the neutron energy distribution of the NEXT UKAL setup has not been fully studied. Using the full NEXT\emph{sim} simulation (GEANT4 interactions and photosensor response), neutron ToFs were measured and the mean of each segment's ToF distribution was plotted against the layer number corresponding to the reconstructed photon CoM, as described in Sect. \ref{sec:nMulti}. In doing so, a position map similar to what is shown in Fig. \ref{fig:PSPMTImage} can be made using simulated data and the same position criteria can be applied to the simulated data as experimental data.

 Fig. \ref{fig:simToFvsSeg} shows the expected prototype position-dependent timing behavior when detecting $\sim$1~MeV neutrons. The same methods used to make cuts on the resistive network event positions in the experimental data were used to determine the column ID from optical photon center-of-mass calculations in the simulated data. The error bars on the data in Fig. \ref{fig:simToFvsSeg} are the statistical uncertainty in the mean for each column. 1.03~MeV neutrons traverse a single cell thickness (6~mm in simulations) in 0.429~ns based on neutron ToF calculations, represented by the dashed red line. The average shift in the mean ToF per column ($\delta_{ToF}$) is equal to the slope of a first order polynomial fit to data. The first row in Table \ref{tab:Slopes} shows the shift in ToF per column, $\delta_{ToF}$=0.415~ns, extracted from a linear fit to the simulated data.
 
$\delta_{ToF}$ from the simulated data agrees with the calculated $\delta_{ToF}$ within 3.2\%, but the data are not well described with a linear fit. This deviation is largely due to neutron multiple-scattering effects, which broaden the photon arrival time distribution and slightly increasing the measured ToF for events with reconstructed positions within the inner segments. In-depth NEXT\emph{sim} simulations will be used to correct for these effects in experimental data. 

\begin{figure}[t]
  \centering
  \includegraphics[width=\linewidth, trim={0 0 0 0}, clip]{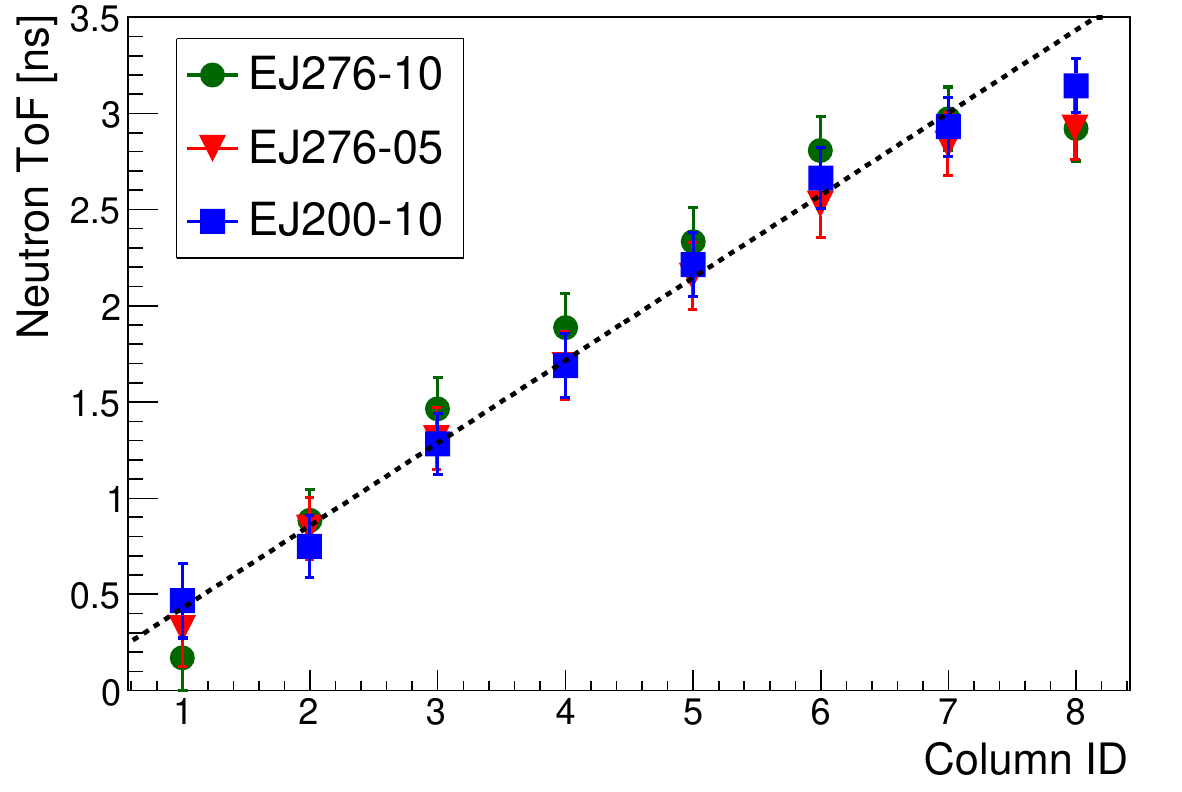}
  \caption{Plots showing the ToF shift per segment for each prototype: EJ276-10 (green), EJ276-05 (red), and EJ200-10 (blue). Data shown correspond to ToF measurements for $\sim$1~MeV neutrons. The data have been shifted appropriately to lie on the same scale. Errors are calculated as the combined uncertainty of the statistical uncertainty in the mean and the timing variations between the four rows in a single column. The black dashed line represents the expected $\delta_{ToF}$ based on ToF calculations for 1.03~MeV neutrons.}
  \label{fig:ToFvsSeg}
\end{figure}

\subsection{Experimental results}

To demonstrate the feasibility of NEXT, analyses were completed for $\sim$1~MeV neutrons detected in three different segmented NEXT prototypes: EJ276-10 (10-inch EJ-276 4$\times$8 array), EJ276-05 (5-inch EJ-276 4$\times$8 array) and EJ200-10 (10-inch EJ-200 4$\times$8 array). For 1.03(3)~MeV neutrons, the mean of the ToF distributions should shift by 0.429~ns for each successive column. By making the appropriate position cuts, the mean ToF for each column was extracted using a Gaussian fit to 1.5$\sigma$ width for the ToF distribution maximum. $\delta_{ToF}$ for each prototype was determined from a first-order polynomial fit to the average ToF vs. column ID data shown in Fig. \ref{fig:ToFvsSeg}.

\begin{table}[t]
\caption{Slopes from first-order polynomial fits to simulated and real ToF data for 1.03(3)~MeV neutrons, yielding $\delta_{ToF}$ in $\left[\sfrac{ns}{col}\right]$ .}
\label{tab:Slopes}
\begin{center}
\begin{tabular}{p{0.2\linewidth} c}
 \hline
 Prototype & $\delta_{ToF}$ $\left[\frac{ns}{col}\right]$  \\
 \hline
 \hline
 NEXT\emph{sim}    & $0.415\pm0.018$  \\
 EJ276-10          & $0.439\pm0.013$  \\
 EJ276-05          & $0.402\pm0.013$  \\
 EJ200-10          & $0.424\pm0.010$  \\
 \hline
\end{tabular}
\end{center}
\end{table}
The $\delta_{ToF}$ values from the fits for each prototype are shown in Table \ref{tab:Slopes}. Overall, each detector exhibited the expected position-dependent timing characteristics, with a clear shift in ToF measurements from column to column and a similar higher-order behavior over the linear fit also evident in the simulations. 
Future publications will contain results from continued analysis on this data set as well as other ToF data acquired for different neutron energies.

\section{Summary}
After extensive development guided by simulations and single-bar tests, a NEXT prototype has been built with 4$\times$8 segmentation. The NEXT concept seeks to improve neutron energy measurements in time-of-flight configurations using simple position-sensitive readouts and economically feasible scintillators with neutron-gamma discrimination capabilities. The prototype meets design goals with a 6~mm layer thickness and 548~ps time-of-flight resolution. The position-dependent timing relationship was validated through analysis of 1.03(3)~MeV neutron ToF measurements using the NEXT prototypes, showing clear shifts in the average time-of-flight between successive scintillation layers. A proof-of-principle neutron energy measurement was made using fission neutrons from a $^{252}$Cf source. The measured neutron yield matched the expected distribution based on GEANT4 prototype simulations, further validating the segment-dependent energy calculations. 
SensL\textsuperscript{\texttrademark} J-Series SiPMs successfully measured gamma-ray times-of-flight with approximately 500~ps timing resolution, validating SiPMs as potential detectors for future development of small-scale ToF arrays.
While NEXT is still in the prototyping phase, it will be continually updated to incorporate further advancements in n-$\gamma$ discrimination for solid-state scintillators and position-sensitive photodetector technologies.\\

\section*{Acknowledgments}
This research was sponsored by the U.S. Department of Energy, National Nuclear Security Administration under the Stewardship Science Academic Alliances program through DOE Award No. DE-NA0002934 and DE-NA0002132 and the U.S. Department of Energy Office of Science through DOE Award No. DE-SC0016988.

\bibliography{NEXT_NIM}

\end{document}